\documentclass[preprintnumbers,article,amsmath,amssymb,floatfix,10pt,prd,onecolumn,superscriptaddress,nofootinbib]{revtex4}

\usepackage{doi}
\usepackage{hyperref}
\hypersetup{
  colorlinks=true,        
  linkcolor=blue,         
  citecolor=cyan,         
}

\usepackage{graphicx}
\usepackage{dcolumn}
\usepackage{bm}
\usepackage{color}
\usepackage{enumitem}
\usepackage{amsmath}
\usepackage{amssymb}

\usepackage{bbm}
\usepackage{amsfonts}
\usepackage{mathrsfs}
\usepackage{latexsym}
\usepackage{epsfig}
\usepackage{epstopdf}
\usepackage{epstopdf}
\usepackage{graphicx}
\usepackage{amssymb}
\usepackage{amsmath}
\usepackage{dcolumn}
\usepackage{bm}
\usepackage{color}
\usepackage{comment}
\usepackage{xcolor}

\begin{document}
\title{\bf Constraining study of circular orbits and accretion disk around nonlinear electrodynamics black hole}

\author{A. Ditta}
\email{adsmeerkhan@gmail.com}\affiliation{Department of mathematics, The Islamia University of Bahawalpur,\\
Bahawalpur-63100, Pakistan}

\author{G. Mustafa}
\email{gmustafa3828@gmail.com}
\affiliation{Department of Physics, Zhejiang Normal University,
Jinhua 321004, China}

\author{G. Abbas}
\email{ghulamabbas@iub.edu.pk}\affiliation{Department of mathematics, The Islamia University of Bahawalpur,\\
Bahawalpur-63100, Pakistan}

\author{Farruh Atamurotov}
\email{atamurotov@yahoo.com}

\affiliation{Inha University in Tashkent, Ziyolilar 9, Tashkent 100170, Uzbekistan}
\affiliation{Akfa University, Milliy Bog' Street 264, Tashkent 111221, Uzbekistan}
\affiliation{National University of Uzbekistan, Tashkent 100174, Uzbekistan}

\author{Kimet Jusufi}
\email{kimet.jusufi@unite.edu.mk}
\affiliation{Physics Department, State University of Tetovo,
Ilinden Street nn, 1200, Tetovo, North Macedonia}

\begin{abstract}
The very latest observation of $M87$ supermassive black hole (BH) by the Event Horizon Telescope (EHT) provides the accretion onto BHs is an interesting study in the theory of gravity. We study the geodesics structure and accretion near a nonlinear electrodynamics BH in strong and weak field approximations. These approximations provide the disc-like structure under the geodesic motion and accretion around the BH. Near the equatorial plane, we provide some new reasons to make circular orbits and accretion of test particles around the BH. Then we investigate perturbations, the critical speed of the fluid and the mass accretion rate of particles around the central object. The physical validity of this study shows that the parameter $\beta$ and $Q$ play an important role in the circular orbits and the mass accretion rate in strong and weak field approximations.
\end{abstract}

\maketitle

\date{\today}

\section{Introduction}

Black holes are quite fascinating objects revealed by the theory of general relativity (GR). In the universe, they are known as robust sources of the gravitational field and are also likely to have a high spin and magnetic intensity. Owing to these features, black holes are the supreme research laboratory for studying both gravity and matter in astrophysical experiments. Currently, some observational evidence has confirmed the presence of BHs. The first was the innovation of gravitational waves brought out from a binary BH merger by LIGO and Virgo collaboration \cite{1}. Another amazing revolutionary is the first image of $M87^{\ast}$ BH shadow \cite{2,3} and the very recently released image of Sgr $A^{\ast}$ \cite{4} by the EHT through a very long baseline interferometry. Furthermore, the perceived electromagnetic spectrum of accretion disks can be responsible for the existence of BHs \cite{5,6,7}. These latest achievements provide an informative response to the appreciation of the theory of GR and the nature of accretion disks near supermassive BHs in the strong gravity regime around the event horizon and may be assumed as a way to analyze modified theories of gravity.

Accretion disks are formed due to the rotation of gaseous materials that travel in bounded orbits around the central mass because of the gravitational force, such as neutron stars, supermassive BHs, and Young Stellar Objects in Active Galactic Nuclei. In such systems, particle orbits are stable, but when the orbits of these materials become unstable, an accretion will occur. It is generally believed that massive central objects such as BHs capture particles from fluid in their vicinity and increase their mass through the accretion process. So, the study of the geodesic motion of particles in the vicinity of BH and particularly the analysis of some characteristic radii such as innermost stable circular orbits ($r_{isco}$) and marginally bound orbits (rmb) are remarkable issues for a suspicious study of the subject matter. These radii are essential in the study of BH accretion disks, where the inner edge of the disk coincides with the innermost stable circular orbit $(ISCO)$ and the efficiency of the energy released \cite{8}.

The locality of stable or unstable circular orbits is consistent with the minimum or maximum of the effective potential correspondingly. In Newtonian theory, the effective potential has a minimum for any value of the angular momentum, and then it has no minimum radius of a stable circular orbit, $(ISCO)$ \cite{9}. But this position is altered when the effective potential has a difficult form liable to the particle angular momentum and other parameters. Therefore, in GR and for the particles moving near the Schwarzschild BH, the effective potential has two extrema for any value of angular momentum. But, only for a particular value of angular momentum do the two points happen together. This point presents $ISCO$ where is placed at $r = 3r_g$ \cite{9,10,11,12,13,14} where $r_g$ is the Schwarzschild radius.

According to the metrics, the effects of spacetime affect the positions of these radii and some parameters such as specific energy, angular momentum, and angular velocity. A lot of research problems are dedicated to studying these radii and their physical structure. The effects of $ISCO$ near the Kerr BH were studied by Ruffini et al. \cite{15} and Bardeen et al. \cite{16}. Even Hobson et al. \cite{17} defined these properties in their textbook on GR. The efficiency of accretion disks, $\eta$, for Schwarzschild and Kerr BHs was calculated by Novikov and Thorne \cite{18}. The Kerr-like metric was raised by Johannsen and Psaltis \cite{19} and then Johannsen \cite{20} formulated the accretion disks near such BHs. The geodesic structure and the circular orbits of charged particles near weakly magnetized rotating BHs are obtained by Tursunov et al. \cite{21}. Since in an accretion disk the particles move in stable orbits but when perturbations (due to restoring forces) act on the particles, oscillations near the circular orbit are produced in radial and vertical directions with epicyclic frequencies. Therefore, the study of orbital and epicyclic frequencies plays an important role in the physics of accretion disks near the BHs. Isper \cite{22, 23}, Wagoner \cite{24}, Kato \cite{11} and Ortega-Rodriges et al. \cite{12} have studied in this field.

The exact and analytic solutions obtained by Bondi \cite{25}, for characterizing different astrophysical scenarios have been involved in the continuous evolution of the accretion theory. Furthermore, analytic solutions are fundamental equipment as benchmark tests for numerical codes \cite{26}. This model was prolonged by Michel \cite{27} to a relativistic regime by assuming a Schwarzschild BH. Then again, analytic solutions to the known wind accretion scene have been developed by Bondi and Hoyle \cite{28}, further Hoyle and Littleton \cite{29} in the Newtonian framework and also by Tejeda and AguayoOrtiz \cite{30} in the relativistic framework of a Schwarzschild BH. Numerous analytic and numerical analyses have further prolonged the work of spherical accretion, e.g. \cite{31,32,33}, and also of wind accretion, e.g. \cite{34,35,36}. Further, Abbas and Ditta obtained the useful properties of accretion for motivation of GR onto a class of BHs \cite{37,38,39,40,41,42,43,44,45}.

So, there is a blank space in the literature because no study has been carried out on accretion disks for nonlinear electrodynamic BH with weak and strong field approximations. However, this study is the motivation of GR and we can honestly fill this blank in the present paper. Our main aspect of this paper is to study the non-rotating BH solutions of an accretion disk in strong and weak fields. For this, we assume an equatorial plane with a polar coordinates system. Firstly, we see the variations in horizons near the geometry of BHs. Then we analyze the circular orbits and effective potential as well. In order to study the locations of circular orbits such as innermost stable circular orbits $r_{isco}$ marginally bound circular orbits $r_{mb}$ and photon sphere $r_{ph}$. The maximum radiation efficiency and binding energy with epicyclic frequencies are also investigated. Finally, the critical accretion with some dynamical parameters of isothermal fluid is obtained.

Recently, Tretyakova \cite{46}, analyzed geodesics in the Horndeski BH observational properties. Further, Salahshoor and Nozari \cite{47} used a similar method to analyze the circular orbits and accretion disks in detail in a class of Horndeski/Galileon BHs. In the present study, we also introduce a general class of nonlinear electrodynamic BH and used a similar procedure to find out the new accretion results for the motivation of GR. The arrangement of this paper is as follows: In section \textbf{II}, we introduce the nonlinear electrodynamic BH spacetime. The wide-range calculations for a test particle motion are formulated in section \textbf{III}. Further, the circular mechanism and oscillations are reviewed in the subsequent sections respectively. In section \textbf{IV}, we found the critical speed of the flow, mass accretion, and its time variation for the perfect fluid. In section \textbf{V}, we analyzed the physical significance of all of these results for strong and weak fields. We summarized the results of the paper in section \textbf{VI} by saying that the parameters $Q$ and $\beta$ have greatly suppressed strong and weak field approximations. Throughout the calculations, we consider the geometric units $G=c=1$, and the spacetime signature $(-,+,+,+)$.

\section{Nonlinear Electrodynamic Black Hole}
The non-rotating nonlinear electrodynamics BH solutions are obtained by the general gravitational action \cite{48,49} which is given as
\begin{equation}
S=\int\sqrt{-g}d^{4} x\left[\frac{R}{2k^2}-\frac{\textsl{\textbf{F}}}{2\beta \textsl{\textbf{F}}+1}\right],\label{1}
\end{equation}
where $g=det(g_{\mu\nu})$, $R$ is a Ricci scalar, $k^{-1}$ is a reduced Plank mass, $\textsl{\textbf{F}}=\frac{1}{4}F_{\mu\nu}F^{\mu\nu}$
and $\beta$ is the dimensionless parameter of nonlinear electrodynamic BH. For a solution, we consider the spherically symmetric spacetime
\begin{equation}
ds^{2}=-f(r)dt^{2}+\frac{1}{f(r)}dr^{2}+r^{2}
(d\theta^{2}+ \sin^{2}\theta d\phi^{2}),\label{2}
\end{equation}
with
\begin{equation}
f(r)=1-\frac{2M}{r}+\frac{Q^2}{r^2}-\frac{C^2k^2}{2r^2}+\frac{C^2k^2}{30r^2}(5\xi^3-22\xi^2+32\xi),\\\label{3}
\end{equation}
where $M, Q$ and $C$ are mass, charge and integration constant, respectively. Also, we have
\begin{equation}\label{4}
\xi=\frac{12\sqrt{3}\sqrt{\beta}r^2-\beta C \lambda^{3/4}}{12\beta C \lambda^{1/4}},
\end{equation}
and
\begin{eqnarray}\label{5}
\lambda&=&\frac{6\sqrt[3]{6}r^2(\sqrt[3]{2}\beta^{2/3}\sqrt[3]{C}\gamma^{2/3}-8\sqrt[3]{3}\beta C)}{\beta^{4/3}C^{5/3}\sqrt[3]{\gamma}},\\
\label{6}
\gamma&=&\sqrt{3}\sqrt{256\beta C^2+27r^4}+9r^2.
\end{eqnarray}
The strong and weak field solutions can be obtained from the metric function $f(r)$ by applying $r\rightarrow 0$ and $r\rightarrow\infty$ respectively,
such that
\begin{eqnarray}\label{7}
f(r)_s&=&1-\frac{2M}{r}+\frac{Q^2}{r^2}-\frac{C^2k^2}{2r^2}+\frac{16C^{3/2}k^2}{15\beta^{1/4}r},\\
\label{8}
f(r)_w&=&1-\frac{2M}{r}+\frac{Q^2}{r^2}-\frac{\beta C^4k^2}{10r^6}.
\end{eqnarray}
\begin{figure}
\centering \epsfig{file=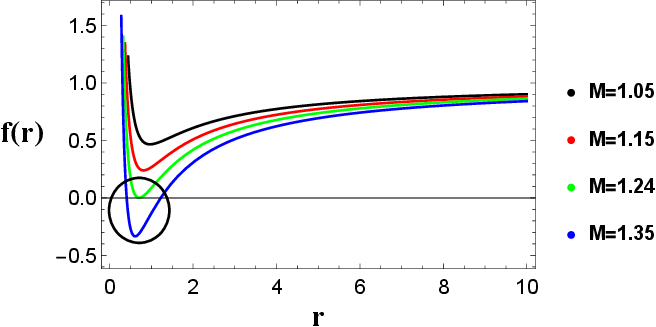, width=.47\linewidth,
height=2.2in}\epsfig{file=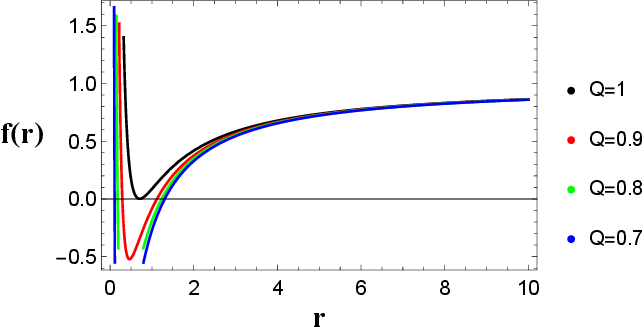, width=.47\linewidth,
height=2.2in}
\centering \epsfig{file=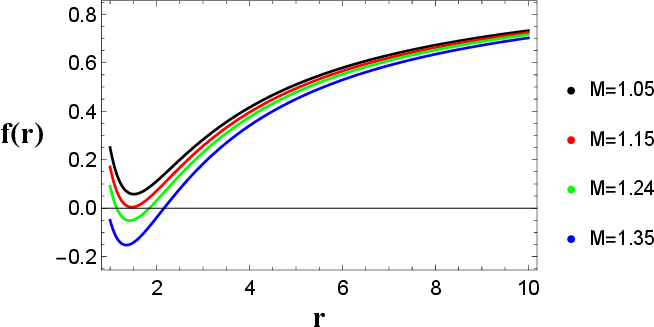, width=.47\linewidth,
height=2.02in}\epsfig{file=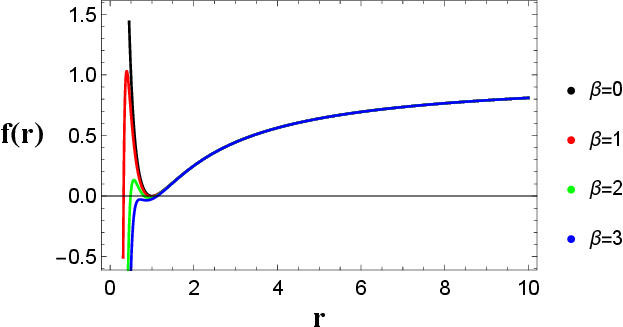, width=.47\linewidth,
height=2.02in}\caption{\label{F1} Variation of horizon of strong and weak fields for $M$, $Q$ and $\beta$.
Upper plots for the strong field while lower plots for the weak field.
This behavior is for the equations $(7, 8)$ and the parameters are
$k=1, C=1$ and different values of $M$, $Q$ and $\beta$. The strong field behavior is depicted for the variation of mass $M$ and the nonlinear electrodynamic parameter $Q$, while the weak field behavior is depicted for the variation of mass $M$ and the nonlinear electrodynamic parameter $\beta$.}
\end{figure}
It is noted that in the strong field limit $f(r)_s$, the effects of nonlinear electrodynamics are highly effective and cannot be detached unless. However, in weak field limit, the last term goes more rapidly as $r\rightarrow\infty$ such that $f(r)_w$ and the metric function $f(r)$ are asymptotically Reissner-Nordstrom and hence the effects of nonlinear electrodynamics can simply be removed by applying $\beta\rightarrow 0$. Fig. (\ref{F1}) has the following key points:
\begin{itemize}
  \item In the upper left plot, the curves (green and blue) pass through the circular disk have an event and two horizons (inner and outer) in a strong field for different values of $M$. An event horizon occurs for $M=1.24$, while inner and outer horizons are appear for $M=1.35$  with $k=1, C=1, Q=1, \beta=1$ .
  \item In the upper right plot, the black curve has an event horizon at $Q=1$ while the red curve has two horizons
  at $Q=0.9$ in a strong field for  $k=1, C=1, M=1.05, \beta=1$.
  \item An event and two horizons are also present in the left bottom plot for the weak field but for different values of mass $M=1.15, 1.24$, as compared to strong field by using $k=1, C=1, Q=1, \beta=1$.
  \item The right bottom plot has an event horizon for $\beta=0$ in a weak field and has no two horizons 
  for different values of $\beta$
with $k=1, C=1, Q=1, M=1.05$.
\end{itemize}
It is noted that in a strong field, for smaller values of $M$, the curves are shifted outward to an event horizon. In the case of a weak field, for larger values of $M$, the curves are shifted inward to an event horizon. Both the left plots have Cauchy horizons in strong and weak fields, while the right plots have not.

\section{General formulism of geodesic motion}
In this section, we formulate the general results for the geodesic motion of test particles by the underlying static and symmetric spacetime. For these results, we consider two Killing vectors $\zeta_t=\partial_t$ and $\zeta_\phi=\partial_\phi$ parallel two constants of motion $E$ and $L$ (conserved energy and angular momentum) by the trajectory as follows.
\begin{eqnarray}\nonumber
E&=&-g_{\mu\nu}\zeta^\mu _t u^\nu\equiv-u_t,\\
\label{9}
L&=&g_{\mu\nu}\zeta^\mu _\phi u^\nu\equiv u_\phi,
\end{eqnarray}
where $u^\mu = \frac{dx^\mu}{d\tau}=(u^t,u^r,u^\theta,u^\phi)$ is the $4$-velocity of the moving particles
and the particles obey the normalization condition $ u^\mu u_\mu = -1$, we obtain
\begin{equation}
g_{rr}(u^r)^2+g_{\theta\theta}(u^\theta)^2+g^{tt}(u_t)^2+g^{\phi\phi}(u_\phi)^2=-1.\label{10}
\end{equation}
Taking $(\theta=\pi/2)$, for the equatorial plane then from Eqs. (\ref{9}) and (\ref{10}), we obtain
\begin{eqnarray}\label{11}
u^t&=&\frac{E}{f(r)},\\\nonumber
u^\theta&=&0,\\\nonumber
u^\phi&=&\frac{L}{r^2},\\\nonumber
u^r&=&\left[f(r)\left(-1+\frac{E^2}{f(r)}-\frac{L^2}{r^2}\right)\right]^{\frac{1}{2}}.
\end{eqnarray}
Now, the conserved energy equation with an effective potential $V_{eff}$ for the motion of the test particle can be written as
\begin{equation}
(u^r)^2+V_{eff}=E^2.\label{12}
\end{equation}
\begin{figure}
\centering \epsfig{file=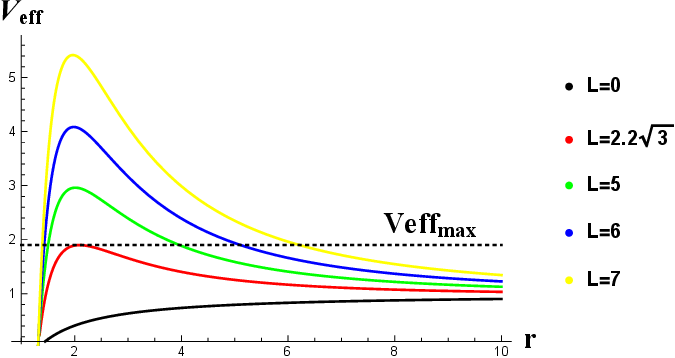, width=.45\linewidth,
height=2.2in}\epsfig{file=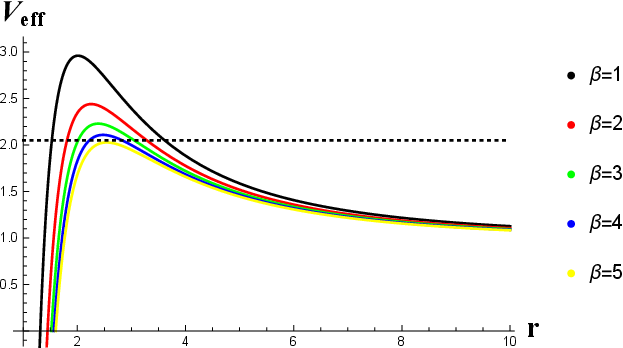, width=.45\linewidth,
height=2.2in}
\centering \epsfig{file=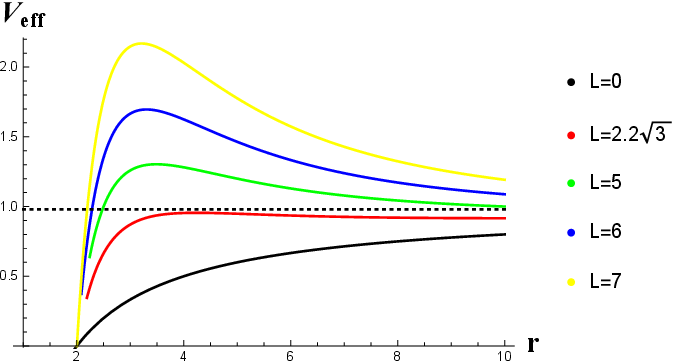, width=.45\linewidth,
height=2.02in}\epsfig{file=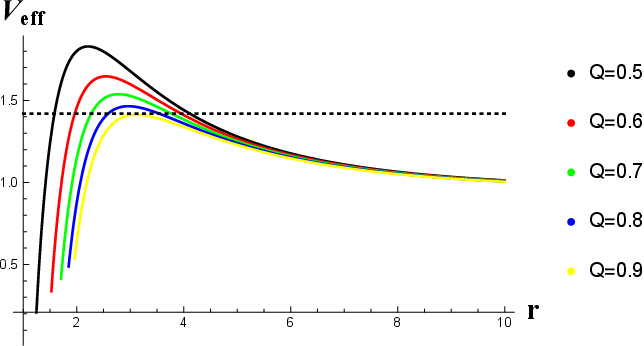, width=.45\linewidth,
height=2.02in}\caption{\label{F2} Variation of $V_{eff}$ of strong and weak fields for $L$, $Q$ and $\beta$.
Upper plots for the strong field while lower plots for the weak field.
This behavior is for the equations (\ref{58}) and (\ref{59}). The strong field behavior is depicted for the variation of effective potential $L$ and the nonlinear electrodynamic parameter $\beta$, while the weak field behavior is depicted for the variation of effective potential $L$ and the nonlinear electrodynamic parameter $Q$ .}
\end{figure}
\begin{equation}
V_{eff}=f(r)\left[1+\frac{L^2}{r^2}\right].\label{13}
\end{equation}
Clearly, the effective potential of particles depends on strong and weak fields parameter $f(r)$ and the angular momentum. The analysis of effective potential is very important in geodesic motion. Since the location of the circular orbits can be found by the local extremum of the effective potential. In Fig. (\ref{F2}), the effective potential has the following key points:
\begin{itemize}
  \item In the left upper plot, there is no extremum for $L=0$ (black curve) in strong field. The first extremum is observed at $V_{eff}=1.95$ for $L=2.2\sqrt{3}$ (red curve) by taking $k=1$, $C=1$, $Q=0.5$, $M=1.05$ and $\beta=1$.
  \item In the right upper plot, the first extremum is observed at $V_{eff}=2.0$ for $\beta=5$ (yellow curve) with $k=1, C=1, Q=0.5, M=1.05$ and $L=5$.
  \item In the left bottom plot, the first extremum is observed at $V_{eff}=1.0$ for $L=2.2\sqrt{3}$ (red curve) by considering $k=1, C=1, Q=0.5, M=1.05$ and $\beta=1$.
   \item In the right bottom plot, the first extremum is observed at $V_{eff}=1.41$ for $Q=0.9$ (red curve) with $k=1, C=1, Q=0.5, M=1.05, \beta=1$ and $L=5$ .
  \item $V_{eff}$ increases for increasing values of $L$ and decreasing values of $\beta$ and $Q$. The maximum potential is seen at $L=7$ and $\beta=1$ and $Q=0.5$.
  \item The curves shifted outward for lager values of $L$ and inward for lager values of $\beta$ and $Q$.
\end{itemize}
It is noted that the location of the innermost stable circular orbits is represented by the point of extremum in a strong field at the distance $r>2$. The potential increases for increasing the angular momentum but, it decreases for increasing the parameter $\beta$ and $Q$. The maximum potential of the strong field is larger than the weak field.
\begin{figure}
\centering \epsfig{file=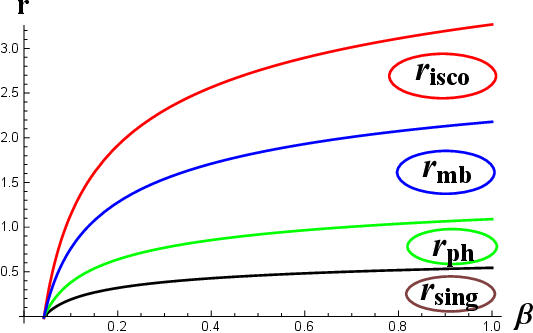, width=.45\linewidth,
height=2.2in}\epsfig{file=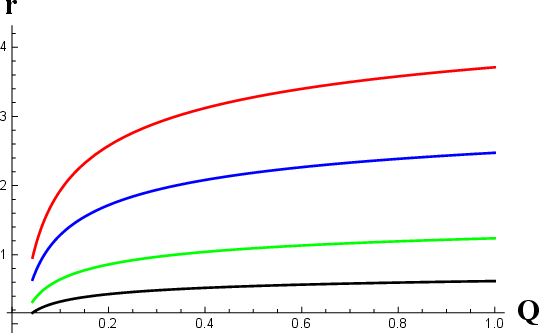, width=.45\linewidth,
height=2.2in}
\centering \epsfig{file=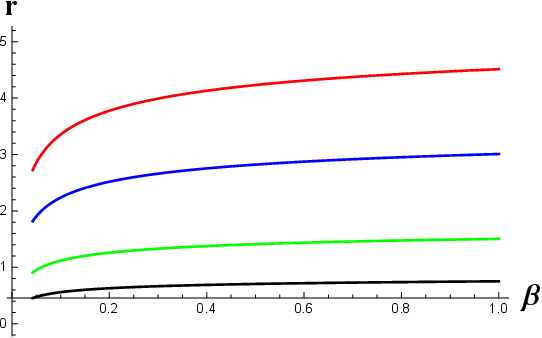, width=.45\linewidth,
height=2.2in}\epsfig{file=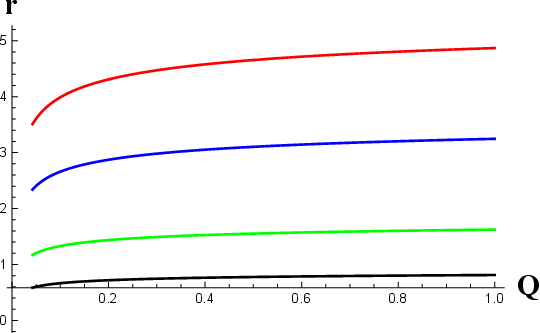, width=.45\linewidth,
height=2.2in} \caption{\label{F3} Variation of the characteristic radii of strong and weak fields w.r.t. $Q$ and $\beta$ for nonlinear electrodynamics BH. This behavior is depicted for the variation of both the nonlinear electrodynamic parameters $Q$ and $\beta$ w.r.t. the radius $r$.}
\end{figure}
The characteristic radii curves versus the radius $r$ and nonlinear electrodynamic parameters $\beta$ and $Q$ for strong and weak fields are plotted in Fig. (\ref{F3}). These radii have the following structure:
\begin{itemize}
  \item The left upper plot shows the effect of the nonlinear electrodynamic parameter $\beta$ on characteristic radius and the comparison between these radii. All the solution curves of strong field shifted inward to the singularity for the fixed values of parameters $M=1, k=1, C=1, Q=0.5$. The location of the characteristic radii depend only on the parameter $\beta$ in the vicinity of the considered BH. For smaller values of $\beta$, the location of $r_{isco}, r_ {ph}, r_ {mb}$ and $r_{sing}$ will be fall very quickly onto the BH.
  \item The right upper plot shows the effect of the nonlinear electrodynamic parameter $Q$ on characteristic radius and the comparison between these radii. All the solution curves of strong field shifted inward to the singularity for the fixed values of parameters $M=1, k=1, C=1, \beta=1$. The location of the characteristic radii depend only on the parameter $Q$ in the vicinity of the BH. For smaller values of $Q$, the location of $r_{isco}, r_ {ph}, r_ {mb}$ and $r_{sing}$ will be closer to the BH.
  \item The left bottom plot shows the effect of the nonlinear electrodynamic parameter $\beta$ on characteristic radius and the comparison between these radii. All the solution curves of weak field shifted inward to the singularity for the fixed values of parameters $M=1, k=1, C=1, Q=0.5$. The location of the characteristic radii depend only on the parameter $\beta$ in the vicinity of the BH. For smaller values of $\beta$, the location of $r_{isco}, r_ {ph}, r_ {mb}$ and $r_{sing}$ will be fall very quickly onto the BH.
  \item The right bottom plot shows the effect of the nonlinear electrodynamic parameter $Q$ on characteristic radius and the comparison between these radii. All the solution curves of weak field shifted inward to the singularity for the fixed values of parameters $M=1, k=1, C=1, \beta=1$. The location of the characteristic radii depend only on the parameter $Q$ in the vicinity of the BH. For smaller values of $Q$, the location of $r_{isco}, r_ {ph}, r_ {mb}$ and $r_{sing}$ will be closer the BH.
  \end{itemize}
We have noted that for smaller values of $Q$ and $\beta$, all the radius falls onto the singularity very quickly. While for large values of $Q$ and $\beta$, the radii shifted outward to the singularity. So, the radius $r_{sing}$ is near to the BH while the radius $r_{isco}$ is away from the BH for strong and weak fields.

\subsection{Circular motions}
Here, we study the circular motion of test particles in an equatorial plane. The radial component for the circular motion $r$ must be constant such that $u^r=\dot{u}^r=0$. So, from Eq. (\ref{12}), we have $V_{eff}=E^2$ and $d/dr V_{eff}=0$. Using these relations, we find specific energy $E$, angular momentum $L$, angular velocity $\Omega_\phi$, and angular momentum $l$ given as.
\begin{eqnarray}\label{14}
E^2&=&\frac{2rf^{2}(r)}{2rf(r)-r^2f'(r)},\\
\label{15}
L^2&=&\frac{r^3 f'(r)}{2rf(r)-r^2f^{'}(r)},\\
\label{16}
\Omega_{\phi}&=&\frac{d\phi}{dt}\equiv\frac{u^\phi}{u^t}\Rightarrow \Omega^2_{\phi}=\frac{1}{2r}f'(r),\\
\label{17}
l^2&=&\frac{L^2}{E^2}=\frac{r^3}{2rf^2(r)}f'(r).
\end{eqnarray}
For being real specific energy and angular momentum, the following condition must be held
\begin{equation}
2rf(r)-r^2f'(r)>0.\label{18}
\end{equation}
Hence, this condition is for the real existence of circular orbits and its limited area can be well-founded by solving the above inequality. The relations $E^2<1$, $E^2=1$ are for the bound and marginally bound orbits hold respectively. From Eq. (\ref{14}), we have the useful equation
\begin{equation}
r^2f'(r)+2rf(r)[f(r)-1]=0.\label{19}
\end{equation}
Generally solving the above equation, one can be easily found the marginally bound orbits. Then from Eqs. (\ref{14}) and (\ref{15}), the energy and momentum will diverge at the radius $r$. So the diverge relation is given by
\begin{equation}
2rf(r)-r^2f'(r)=0.\label{20}
\end{equation}
This relation is very important for the characteristic of the photon sphere. Since the photon moves in a photon sphere on circular orbits. Therefore, the comparison of strong and weak fields plays an important role in the significance of circular orbits.
\begin{figure}
\centering \epsfig{file=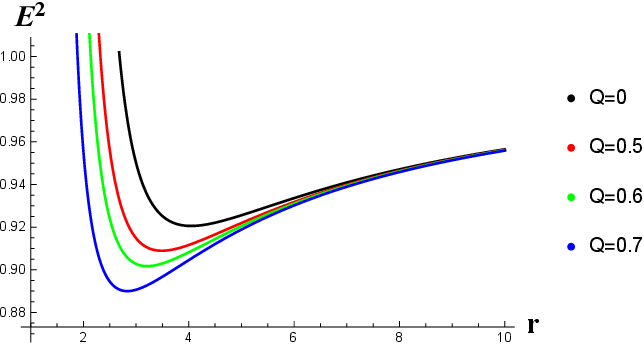, width=.45\linewidth,
height=2.2in}\epsfig{file=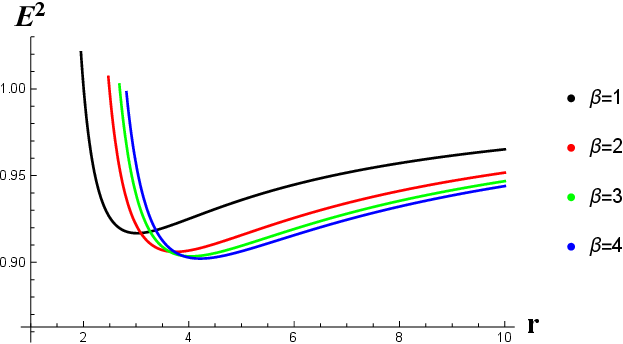, width=.45\linewidth,
height=2.2in}
\centering \epsfig{file=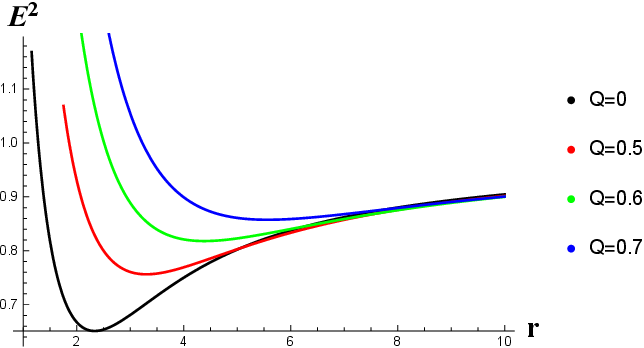, width=.45\linewidth,
height=2.02in}\epsfig{file=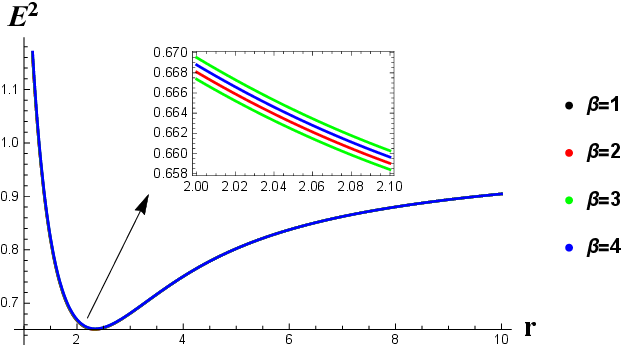, width=.45\linewidth,
height=2.02in}\caption{\label{F4} Variation of energy of strong and weak fields for $Q$ and $\beta$.
Upper plots are for the strong field while lower plots are for the weak field.
This behavior is for the equation (\ref{64}) and (\ref{65}).}
\end{figure}
The specific energy variation is represented in Fig. (\ref{F4})and has the following key points:
\begin{itemize}
  \item Increases the specific energy and decreases the bound orbit radius for smaller values of $\beta$ and larger values of $Q$ in a strong field.
  \item decreases the specific energy and very small change in the bound orbit radius for smaller values of $\beta$ and larger values of $Q$ in a weak field.
  \item We plotted the left plots for taking fix value of $\beta=1$ and variation in $Q$.
  \item We plotted the right plots for taking fix value of $Q=0.5$ and variation in $\beta$.
\end{itemize}
\begin{figure}
\centering \epsfig{file=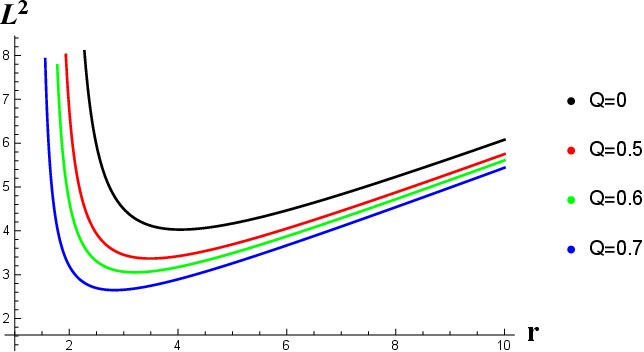, width=.45\linewidth,
height=2.2in}\epsfig{file=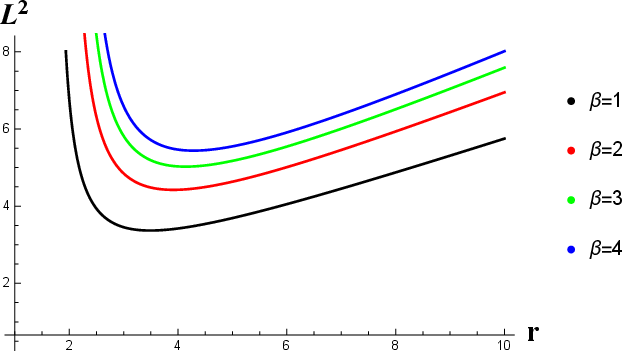, width=.45\linewidth,
height=2.2in}
\centering \epsfig{file=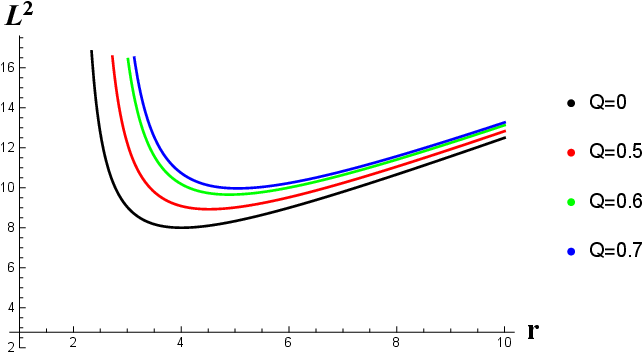, width=.45\linewidth,
height=2.02in}\epsfig{file=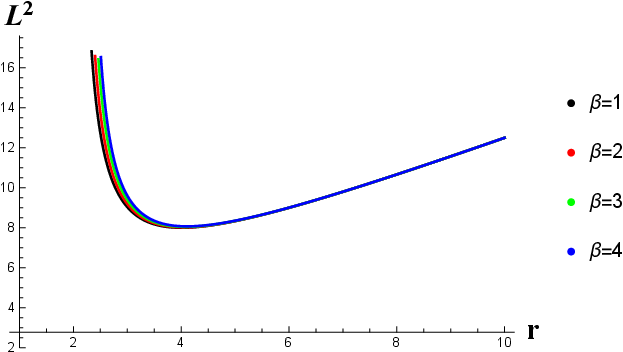, width=.45\linewidth,
height=2.02in}\caption{\label{F5} Variation of angular momentum of strong and weak fields for $Q$ and $\beta$.
Left plots for the strong field while right plots for the weak field.
This behavior is for the equations (\ref{66}), (\ref{67}).}
\end{figure}
The angular momentum variation is represented in Fig. (\ref{F5}) and has the following key points:
\begin{itemize}
  \item The angular momentum decreases for the increasing values of $Q$ and increases for the increasing values of $\beta$ in strong field.
  \item The angular momentum increases for the increasing values of $Q$ and increases for the decreasing values of $\beta$ in weak field.
  \item We plotted the left plots for taking fix value of $\beta=1$ and variation in $Q$.
  \item We plotted the right plots for taking fix value of $Q=0.5$ and variation in $\beta$.
\end{itemize}

\subsection{Stable circular orbits and radiation energy flux}
Stable circular orbits depend on the local minima of the effective potential that is found by the relations
$\frac{d^2}{dr^2}V_{eff}>0$ and for marginally stable circular orbits $\frac{d^2}{dr^2}V_{eff}=0$.
Then from Eq. (\ref{13}), we get
\begin{equation}
 \frac{d^2}{dr^2}V_{eff}=f''(r)(1+\frac{L^2}{r^2})-4f'(r)\frac{L^2}{r^3}+6f(r)\frac{L^2}{(r)^4}.\label{21}
\end{equation}
The accretion possibly corresponds to the characteristic $r < r_{isco}$.
It is noted that the falling materials from rest at infinity accrete onto the BH,
the released gravitational energy of the falling materials may change
into radiation and this energy is the cause of the most energetic astrophysical phenomena.
The radiation energy flux over the accretion disk is due to the radiant energy corresponding to the
specific energy $E$, angular momentum $L$ and angular velocity $\Omega_\phi$
studied by Kato et al. \cite{10}. Then we have
\begin{equation}
K=-\frac{\dot{M}\Omega_{\phi,r}}{4\pi \sqrt{-g}(E-L\Omega_\phi)^2}\int(E-L\Omega_\phi)L_{,r}dr,\label{22}
\end{equation}
where $K$ is a radiation flux, $\Omega_{\phi,r}=\frac{d\Omega_\phi}{dr}$,
$\dot{M}$ is an accretion rate and $g$ is the parameter
which is given by
\begin{equation}
g=det(g_{\mu\nu})=-r^4 \sin^2 \theta.\label{23}
\end{equation}
As our work restrictions are in the equatorial plane, so $\sin\theta=1$.
From the Eqs (\ref{13})-(\ref{16}), we have
\begin{eqnarray}\label{24}
K(r)=&&-\frac{\dot{M}}{4\pi r^4}\frac{r}{\sqrt{2f'(r)}}
\\\nonumber&&\times \frac{[2f(r)-r f'(r)][r f''(r)-f'(r)]}{[2f(r)+r f'(r)]^2}
\int^{r}_{mb}Z(r)dr,
\end{eqnarray}
where
\begin{eqnarray}\label{25}
Z(r)=&&\sqrt{\frac{r}{2f'(r)}}\frac{[2f(r)+r f'(r)][-f''(r)r f(r)+2r f'^2(r)-3f'(r)f(r)]}{[2f(r)-r f'(r)]^2}.
\end{eqnarray}
As the steady-state accretion disk is considered in thermodynamical equilibrium,
then the radiation emitted from the disk in the form of black body radiation.
So, we supposed the relation $K(r)=\sigma T^4(r)$ among the radiation and temperature. Here, $\sigma$ is Stefan's constant.
The detailed study of this relation was explained by \cite{51} and further debated by \cite{52}.
\begin{figure}
\centering \epsfig{file=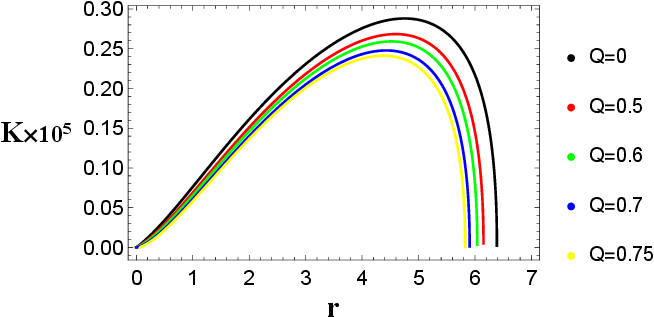, width=.45\linewidth,
height=2.2in}\epsfig{file=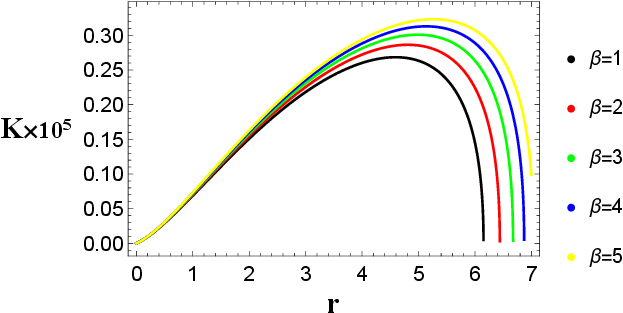, width=.45\linewidth,
height=2.2in}
\centering \epsfig{file=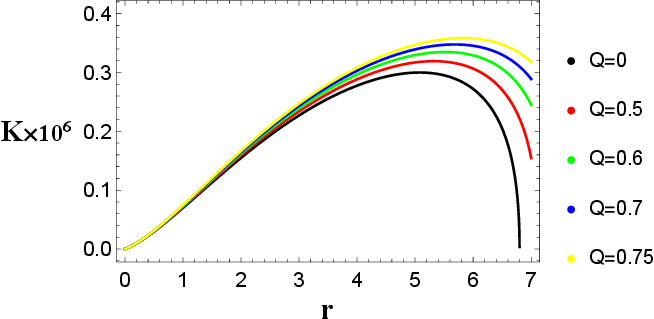, width=.45\linewidth,
height=2.02in}\epsfig{file=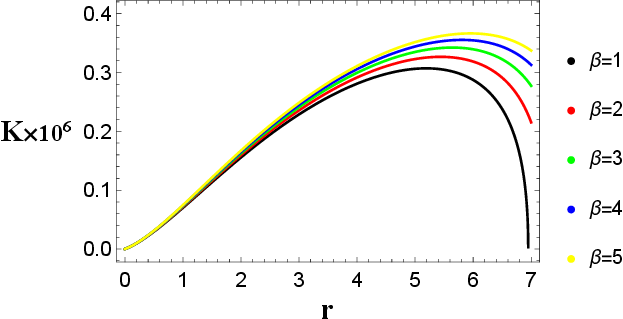, width=.45\linewidth,
height=2.02in}\caption{\label{F6} Variation of emission rate depend on radius $r$ and non linear electrodynamic parameters $Q$ and $\beta$ in strong and weak fields.
The black curves show the picture for no charge that is $Q=0$ and $\beta=1$ in strong and weak fields.}
\end{figure}
The radiation flux is represented in Fig. (\ref{F6}) and has the following key points:
\begin{itemize}
  \item The radiation flux has a maximum for $Q=0$ and it is decreased for increasing values of charge in the left plot of a strong field.
  \item In the right plot of the strong field, the flux is increased for increasing values of $\beta$.
  \item The radiation flux has a minimum for $Q=0$ and it is increased for some different values of charge in the left plot of the weak field.
  \item The same picture is seen in the right plot for $\beta=1$ and some increasing values of $\beta$.
\end{itemize}
The temperature behavior can be seen by using the relation $K=\sigma T^4$.
\begin{figure}
\centering \epsfig{file=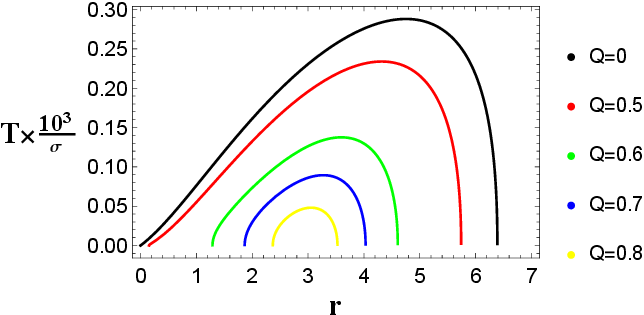, width=.45\linewidth,
height=2.2in}\epsfig{file=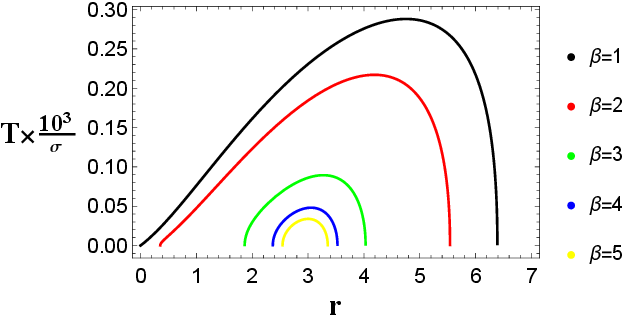, width=.45\linewidth,
height=2.2in}
\centering \epsfig{file=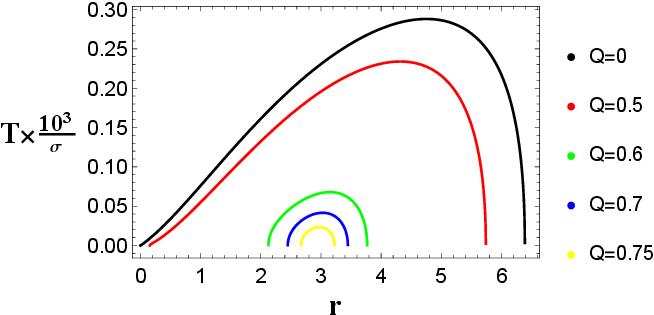, width=.45\linewidth,
height=2.02in}\epsfig{file=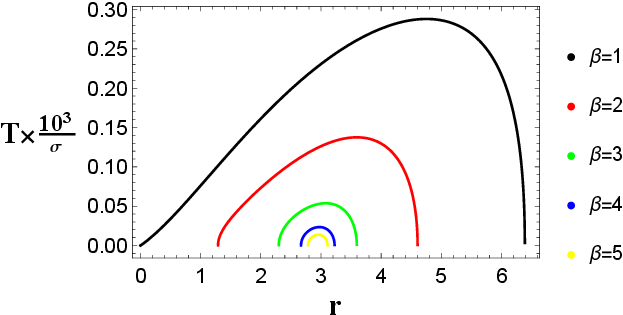, width=.45\linewidth,
height=2.02in}\caption{\label{F7} Variation of temperature depend on radius $r$ and non linear electrodynamic parameters $Q$ and $\beta$ in strong and weak fields.
The black curves show the picture for no charge that is $Q=0$.}
\end{figure}
The temperature profile is represented in Fig. (\ref{F7}) and has the following key points:
\begin{itemize}
  \item The temperature attains a maximum position for $Q=0, \beta=1$ and it is decreased for increasing values of $Q$ and $\beta$ in strong field.
  \item The three curves (green, blue, and Yellow) are close to the bound radius between $(1, 5)$.
  \item The same behavior is seen in the weak field but only the difference is that these three curves
  (green, blue and Yellow) are close to the bound radius between $(2, 4)$ for the same values of the parameters.
\end{itemize}
The efficiency of the accreting fluid is another important analysis of the accretion disk.
The maximum accreting efficiency of transforming energy into the radiative flux
of particles between $ISCO$ and infinity
is defined as the ratio of the binding energy of
the $ISCO$ to the rest of mass energy.
Therefore, the relations of efficiency and maximum accreting efficiency are
$\eta=1-E$ and $\eta^{\ast}=1-E_{ISCO}$ respectively.
\begin{figure}
\centering \epsfig{file=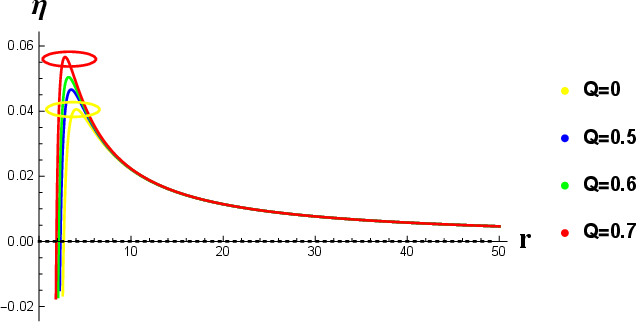, width=.45\linewidth,
height=2.2in}\epsfig{file=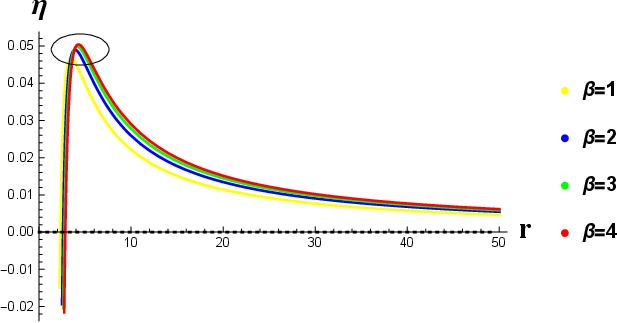, width=.45\linewidth,
height=2.2in}
\centering \epsfig{file=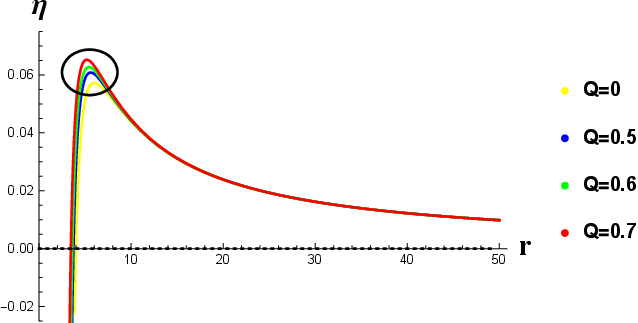, width=.45\linewidth,
height=2.02in}\epsfig{file=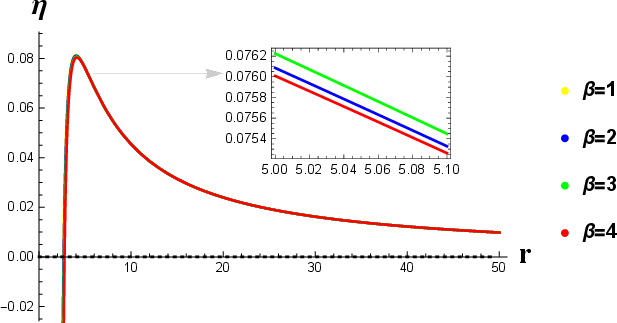, width=.45\linewidth,
height=2.02in}\caption{\label{F8} Variation of efficiency of strong and weak fields for $Q$ and $\beta$.
Left plots for the strong field while right plots for the weak field.
The circular disks denote the maximum efficiency.
This behavior is for the equation $1-E$, where $E$ is taken from (\ref{64}) and (\ref{65}).}
\end{figure}
The efficiency behavior is represented in Fig. (\ref{F8}) and have the following key points:
\begin{itemize}
  \item The efficiency decreases for decreasing values of $Q$ and increasing values of $\beta$ and vice versa in the unstable orbits and the stable orbits are inward to the singularity in strong and weak fields.
  \item Radial frequency denotes by yellow curves while the other curves are for the vertical frequency.
  \item We plotted the left plots for taking fix value of $\beta=1$ and variation in $Q$.
  \item We plotted the right plots for taking fix value of $Q=0.5$ and variation in $\beta$.
\end{itemize}

\subsection{Oscillations}
In the accretion process, various types of Oscillatory motions due to restoring forces are expected.
Restoring forces perform the horizontal and vertical oscillations of perturbations
in accretion disks. These two oscillations are produced by the rotation of
the accretion disk. In this rotation, the equilibrium position is seen
when a particle is transformed to the radial direction.
Further, the accretion disk is also composed of the centrifugal force
as a result of the gravitational field of the central mass.
When the former take over the latter then the fluid element will be pushed outward
or inward to come back to its original position because of the epicyclic frequency $\Omega_r$.
The fluid element produced the harmonic oscillation with vertical
frequency $\Omega_\theta$ by the restoring force in the equatorial plane
(the plane where the element of the fluid is perturbed
in a vertical direction) \cite{10}.
Therefore, three frequencies are noted in a general discussion
for the motion of the fluid element in the accretion disk.
Harmonic vertical motion with vertical frequency $\{\Omega_\theta$, the harmonic radial motion
with radial frequency $\Omega_r$ and the circular motion with orbital frequency $\Omega_\phi\}$.
In this section, we formulate the radial and vertical motions
around the circular equatorial plane.

For the radial and vertical motions, we take the relations $\frac{1}{2}\left(\frac{dr}{dt}\right)^2=V^{(r)}_ {eff}$
and $\frac{1}{2}\left(\frac{d\theta}{dt}\right)^2=V^{(\theta)}_ {eff}$ respectively.
From Eq. (\ref{9}), we put $u^\theta=0$ for the radial motion, and for vertical motion, we put $u^r=0$.
By setting $u^r=\frac{dr}{d\tau}=\frac{dr}{st}u^t$ and $u^\theta=\frac{d\theta}{d\tau}=\frac{d\theta}{st}u^t$, we have
\begin{eqnarray}\label{26}
\frac{1}{2}\left(\frac{dr}{dt}\right)^2&=&-\frac{1}{2}\frac{f^3 (r)}{E^2}\left[1+\frac{E^2}{f(r)}+\frac{L^2}{r^2\sin^2 \theta}\right]=V^{(r)}_ {eff}.\\\nonumber
\frac{1}{2}\left(\frac{d\theta}{dt}\right)^2&=&-\frac{1}{2}\frac{f^2(r)}{r^2E^2}\left[1+\frac{E^2}{f(r)}+\frac{L^2}{r^2\sin^2 \theta}\right]=V^{(\theta)}_ {eff}.
\end{eqnarray}
For more explanation of radial and vertical epicyclic frequencies around the circular orbit,
some perturbations $\delta r$ and $\delta\theta$ are required. So, taking the time derivative
of radial Eq. (\ref{26}), we get
\begin{equation}
\frac{d^2 r}{dt^2}=\frac{dV^{(r)}_ {eff}}{dr}.\label{27}
\end{equation}
The deviation $\delta r=r-r_0$ is for the perturbed particle from its original radius
$r=r_0$ and it has the following perturbed equation of motion
\begin{equation}
\frac{d^2 \delta r}{dt^2}=\frac{dV^{(r)}_ {eff}}{dr}\delta r\Rightarrow \delta \ddot{r}+\Omega^2 _{r}\delta r=0,\label{28}
\end{equation}
the double dots represent the differential with time $t$ and $\Omega^2 (r)=-\frac{d^2}{d\theta^2}V^{(r)}_ {eff}$
whereas applying the same method for vertical direction with deviation $\delta\theta=\theta-\theta_0$, we obtain
\begin{equation}
\frac{d^2 \delta \theta}{dt^2}=\frac{dV^{(\theta)}_ {eff}}{dr}\delta \theta\Rightarrow \delta \ddot{\theta}+\Omega^2 _{\theta}\delta \theta=0,\label{29}
\end{equation}
where $\Omega^2 (\theta)=-\frac{d^2}{d\theta^2}V^{\theta}_ {eff}$. Then from Eq. (\ref{26}),
we have
\begin{equation}
\Omega^{2}_{\theta}=\frac{f^2 (r)L^2}{r^4 E^2}.\label{30}
\end{equation}
and
\begin{eqnarray}\label{31}
\Omega^{2}_{r}=&&\frac{1}{2r^4 E^2}[(L^2+r^2)r^2f^2(r)-f(r)r^4E^2]f''(r)\\\nonumber
&&+2[(L^2+r^2)f(r)-r^2E^2]r^2f(r)f''(r)+2r^2f(r)f'^2(r)(L^2+r^2)\\\nonumber
&&-2r\left[[(L^2+r^2)2f(r)+r^2E^2]r f'(r)+4L^2f^2(r)\right]f'(r)\\\nonumber
&&-4L^2f^2(r)(-\frac{3}{2}f(r)+r f'(r))].
\end{eqnarray}

\section{Basic dynamical equations}
In this section, we find some basic calculations of accretion which were found by
Babichev et al. \cite{52,53}. For these calculations, we take the
Energy-momentum tensor of a perfect fluid
\begin{equation}
T^{\mu\nu} =(\rho+p)u^\mu u^\nu + pg^ {\mu \nu},\label{32}
\end{equation}
where $\rho$ and $p$ are energy density and pressure of accreting fluid whereas
$u^\mu$ is the four-velocity of the fluid.
In an equatorial plane $\theta=\pi/2$, the general form of the
four-velocity is given by
\begin{equation}
u^\mu = \frac{dx^\mu}{d\tau}=(u^t,u^r,0,0),\label{33}
\end{equation}
where $\tau$ denotes the proper time of the geodesic motion of the particles.
The steady-state and spherically symmetric flow obey the
normalization condition $ u^\mu u_\mu = -1$, we obtain
\begin{equation}
u^t =\frac{\sqrt{f(r)+(u^r)^2 }}{f(r)},\label{34}
\end{equation}
due to square root, $u^t$ and $u^r$ may be positive or negative which specifies
the forward or backward time conditions.
As for $u^r<0$ (inward flow), the accretion occurs and the velocity of the fluid is negative
whereas for $u^r>0$ (outward flow) and velocity of the fluid is positive.
Therefore, the conservation laws of energy and momentum
are necessary for the accretion analysis.
Energy conservation is given by
\begin{equation}
T^{\mu\nu}_{;\mu} =0\Rightarrow T^{\mu\nu}_{;\mu}=\frac{1}{\sqrt{-g}}(\sqrt{-g}T^{\mu\nu})_{,\mu}+\Gamma^{\nu}_{\alpha\mu}T^{\alpha\mu}=0,\label{35}
\end{equation}
where $;$ shows the covariant derivative whereas $\sqrt{-g}=r^2\sin^2 \theta$
and $\Gamma$ is the second kind of Christoffel symbol. By simplifications, we obtain
\begin{equation}
r^2u^r(\rho+p)\sqrt{f(r)+(u^r)^2}=N_0,\label{36}
\end{equation}
where $N_0$ represents the integration constant.
The relation between conservation law and four-velocity
via $u_{\mu}T^{\mu\nu}_{;\nu}=0$, we calculate
\begin{equation}
(\rho+p)_{;\nu}u_{\mu}u^{\mu}u^{\nu}+(\rho+p)u^{\mu}_{;\nu}u_{\mu}u^{\nu}+(\rho+p)u_{\mu}u^{\mu}u^{\nu}_{;\nu}+p_{,\nu}g^{\mu\nu}u_{\mu}+p u_{\mu}g^{\mu\nu}_{;\nu}=0.\label{37}
\end{equation}
By the conditions $ u^\mu u_\mu = -1$ and $g^{\mu\nu}_{;\nu}=0$, the above equation reduces to
\begin{equation}
(\rho+p)u^{\nu}_{;\nu}+u^{\nu}\rho_{\nu}=0.\label{38}
\end{equation}
Taking the non-zero components, we obtain
\begin{equation}
\frac{\rho'}{\rho+p}+\frac{u'}{u}+\frac{2r}{r^2}=0.\label{39}
\end{equation}
By integration
\begin{equation}
r^2u^{r}\exp\int\frac{d\rho}{\rho+p}=-N_{1},\label{40}
\end{equation}
where $N_{1}$ represents an integration constant.
Using $u^r<0$, the minus sign is taken on the right-hand side, so we get
\begin{equation}
(\rho+p)\sqrt{\left[(u^r)^2+f(r)\right]}\exp\left(-\int\frac{d\rho}{\rho+p}\right)=N_{2},\label{41}
\end{equation}
where $N_{2}$ represents an integration constant.
By using the above setup, the mass flux is given by
\begin{equation}
(\rho u^\mu)_{;\mu}\equiv\frac{1}{\sqrt{-g}}(\sqrt{-g}\rho u^\mu)_{,\mu}=0,\label{42}
\end{equation}
and also it can be written as
\begin{equation}
\frac{1}{\sqrt{-g}}(\sqrt{-g}\rho u^\mu)_{,r}+\frac{1}{\sqrt{-g}}(\sqrt{-g}\rho u^\mu)_{,\theta}=0.\label{43}
\end{equation}
Therefore the conservation mass equation is given by
\begin{equation}
r^2\rho u^r=N_{3},\label{44}
\end{equation}
where $N_{3}$ is constant of integration.

\subsection{Dynamical parameters}
To continue further, we take isothermal fluid with the equation of state $p=\omega\rho$ whereas $\omega$ is the state parameter.
The flow must be flowing at a constant temperature in these fluids.
Throughout the accretion, the sound speed remains constant for such fluids $p\propto\rho$.
Then, from Eqs. (\ref{40}), (\ref{41}) and (\ref{44}), we have
\begin{equation}
\left(\frac{\rho+p}{\rho}\right)\sqrt{\left[(u^r)^2+f(r)\right]}\exp\left(-\int\frac{d\rho}{\rho+p}\right)=N_{4},\label{45}
\end{equation}
where $N_{4}$ is the integration constant. Using $p=\omega\rho$ in above equation, we get
\begin{equation}
u(r)=\left(\frac{1}{\omega+1}\right)\sqrt{\frac{(N_{4})^2}{f(r)}-(\omega+1)^2}.\label{46}
\end{equation}
Therefore, the radial velocity of strong and weak fields is given by
\begin{eqnarray}\label{47}
u(r)_s&=&\left(\frac{1}{\omega+1}\right)\sqrt{\frac{(N_{4})^2}{\left(1-\frac{2M}{r}+\frac{Q^2}{r^2}-\frac{C^2\omega^2}{2r^2}+\frac{16C^{3/2}k^2}{15\beta^{1/4}r}\right)}
-(\omega+1)^2}.\\
\label{48}
u(r)_w&=&\left(\frac{1}{\omega+1}\right)\sqrt{\frac{(N_{4})^2}{\left(1-\frac{2M}{r}+\frac{Q^2}{r^2}-\frac{\beta C^4k^2}{10r^6}\right)}
-(\omega+1)^2}.
\end{eqnarray}
From Eq. (\ref{44}), we obtain the density of the fluid, given by
\begin{equation}\label{49}
\rho(r)=\frac{N_{3}}{r^2}\frac{(\omega+1)}{\sqrt{{\frac{(N_{4})^2}{f(r)}-(\omega+1)^2}}}.
\end{equation}
Therefore, the energy density of strong and weak fields is given by
\begin{eqnarray}\label{50}
\rho(r)_s&=&\frac{N_{3}}{r^2}\frac{(\omega+1)}{\sqrt{{\frac{(N_{4})^2}{\left(1-\frac{2M}{r}+\frac{Q^2}{r^2}-\frac{C^2\omega^2}{2r^2}+\frac{16C^{3/2}k^2}
{15\beta^{1/4}r}\right)}-(\omega+1)^2}}}.\\
\label{51}
\rho(r)_w&=&\frac{N_{3}}{r^2}\frac{(\omega+1)}{\sqrt{{\frac{(N_{4})^2}{\left(1-\frac{2M}{r}+\frac{Q^2}{r^2}-\frac{\beta C^4k^2}{10r^6}\right)}
-(\omega+1)^2}}}.
\end{eqnarray}

\begin{figure}
\centering \epsfig{file=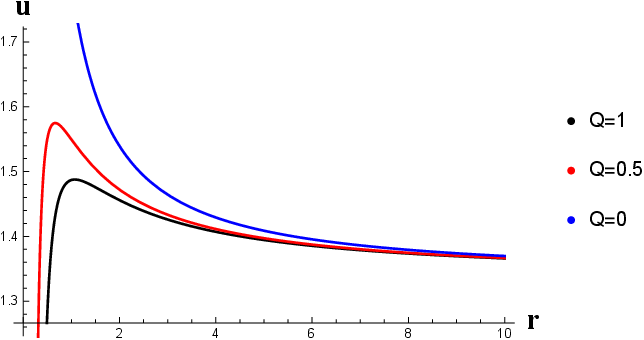, width=.45\linewidth,
height=2.2in}\epsfig{file=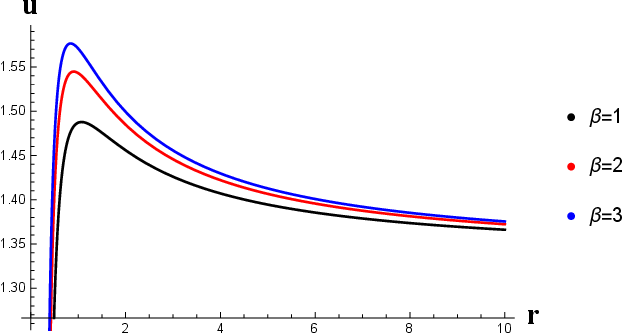, width=.45\linewidth,
height=2.2in}
\centering \epsfig{file=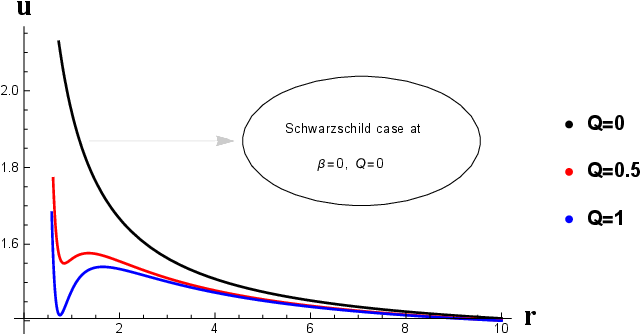, width=.45\linewidth,
height=2.02in}\epsfig{file=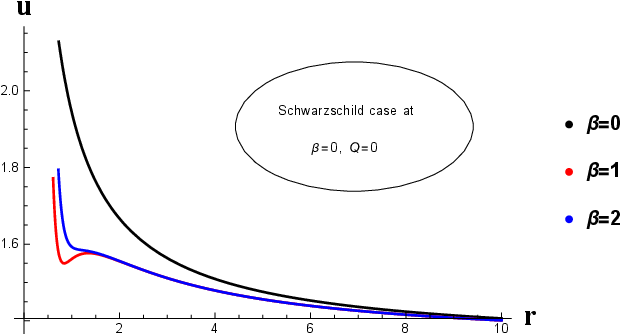, width=.45\linewidth,
height=2.02in}\caption{\label{F9} Variation of velocity of strong and weak fields for $Q$ and $\beta$,
also the equation of sate parameter $k=0.5$.}
\end{figure}
\begin{figure}
\centering \epsfig{file=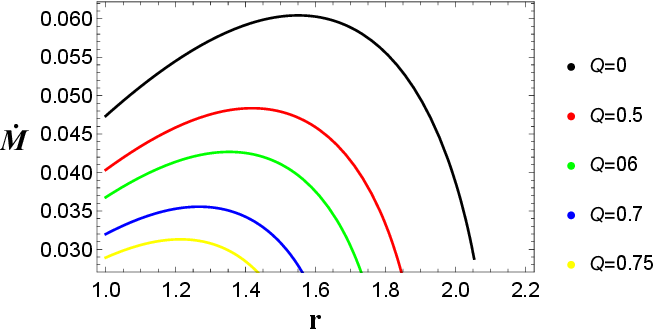, width=.45\linewidth,
height=2.2in}\epsfig{file=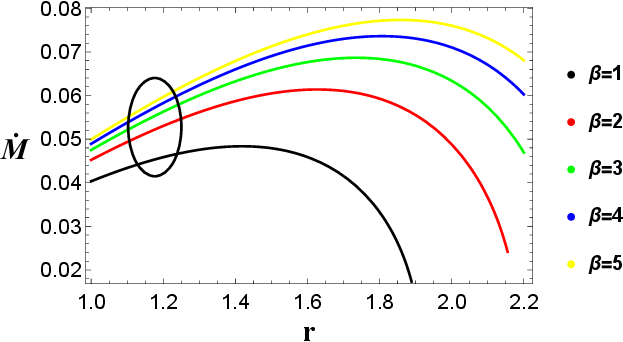, width=.45\linewidth,
height=2.2in}
\centering \epsfig{file=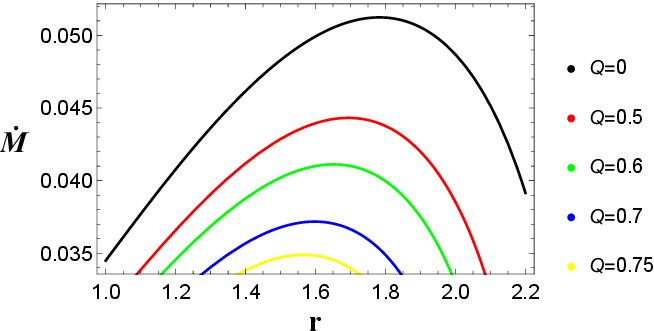, width=.45\linewidth,
height=2.02in}\epsfig{file=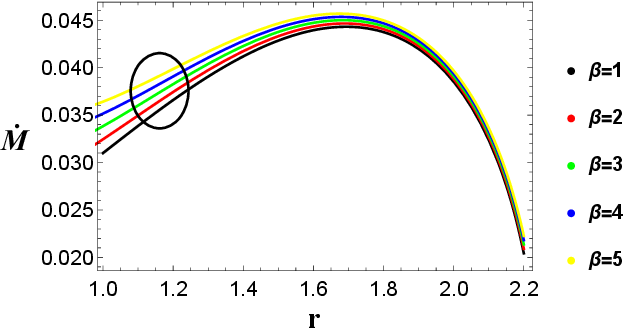, width=.45\linewidth,
height=2.02in}\caption{\label{F11} Variation of the mass accretion rate of strong and weak fields
w.r.t. $Q$ and $\beta$ for nonlinear electrodynamic BH.}
\end{figure}

The velocity variation is represented in Fig. (\ref{F9}) and has the following key points:
\begin{itemize}
  \item The fluid velocity increases by decreasing the charge values ($Q=1, 0.5, 0$), and decreases the bound radius in a strong field.
  \item The increasing behavior is seen for increasing values of ($\beta=1, 2, 3$) in strong field.
  \item The Schwarzschild case (black curve) is seen for taking $\beta=0$ and $Q=0$ only in weak field. We see the different values of parameters affect the radial velocity only in nonlinear case. The velocity of moving particle increases for decreasing values of the parameters. While, in the Schwarzschild case, the value of the parameters is fixed and the velocity of fluid is larger than the nonlinear case.
  \item The blue and red plots show the nonlinear effects of fluid velocity for taking different values of $Q$ and $\beta$ in a weak field.
\end{itemize}
\begin{figure}
\centering \epsfig{file=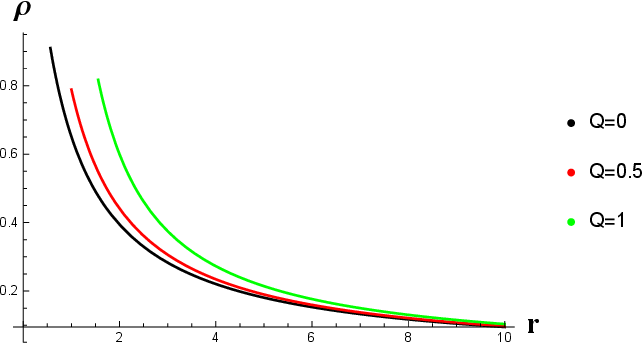, width=.45\linewidth,
height=2.2in}\epsfig{file=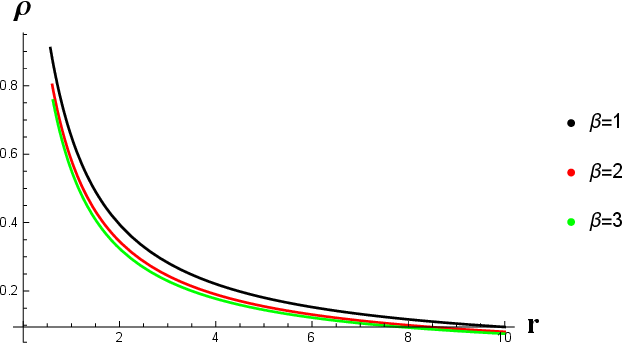, width=.45\linewidth,
height=2.2in}
\centering \epsfig{file=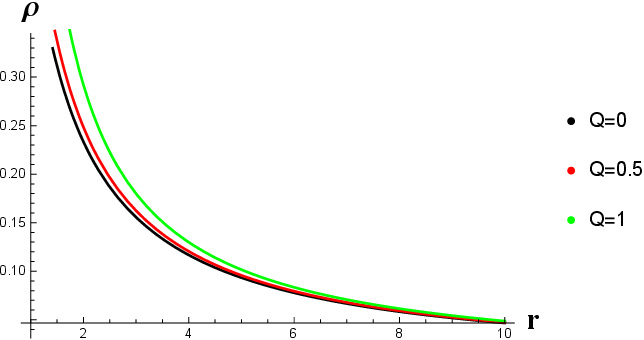, width=.45\linewidth,
height=2.02in}\epsfig{file=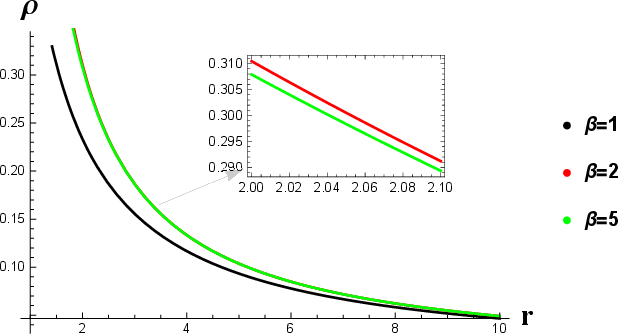, width=.45\linewidth,
height=2.02in}\caption{\label{F10} Variation of density of fluid of strong and weak fields for $Q$ and $\beta$,
also the equation of sate parameter $k=0.5$.}
\end{figure}
Fig. (\ref{F10}) represents the density variation and has the following key points:
\begin{itemize}
  \item The fluid density decreases by increasing the charge values ($Q=0, 0.5, 1.0$),
  and curves shifted outward to the bound radius in a strong field.
  \item The density decreases by increasing the values of ($\beta=1, 2, 3$) and curves shifted inward to the bound radius in strong field.
  \item In weak field, the density increases for increasing $Q$ and $\beta$.
\end{itemize}
The mass accretion rate for the strong and weak fields are plotted in Fig. (\ref{F11}),
therefore the plots have the following structure:
\begin{itemize}
  \item The maximum accretion rate of strong field occurs for no charge ($Q=0$) and ($\beta=5$) between the distance $1.0, 2.05$ (black curve).
  \item The minimum accretion rate of strong field occurs for charge ($Q=0.75$) and ($\beta=1$) between the distance $1.0, 1.4$ (yellow curve).
  \item The maximum accretion rate of weak field occurs for no charge ($Q=0$) and ($\beta=5$) between the distance $1.0, 2.2$ (black curve).
  \item The minimum accretion rate of weak field occurs for charge ($Q=0.75$) and ($\beta=5$) between the distance $1.4, 1.75$ (yellow curve).
  \end{itemize}
In the strong and weak fields, we have noted that for large values of $Q$, the mass accretion rate decreases and the curves are inward to the smaller radii. While for small values of $Q$, the mass accretion rate increases, and the curves are outward to the larger radii. The small variation of mass accretion rate occurs all the solution curves pass through the circular disk for the different values of $\beta$.

\subsection{Mass expansion}
The rate of change in the mass of BH is called mass accretion rate. It is the area times flux around the boundary of BH.
Basically, we measure the mass of BH per unit time. In this case, the accretion rate depends on
the nature of the accreting fluid and also the metric parameters.
The mass of the BH is non-static for quintessence in astrophysical cases.
Due to some processes such as accretion onto the BH and Hawking
radiation, the mass will be changed slowly. The rate of
change of accretion mass can be achieved by integrating the flux of
fluid over the locality of BH and it is denoted by $\dot{M}$.
Therefore, it is given by
\begin{equation}
\dot{M}=-4\pi r^2 u^r(\rho+p)\sqrt{f(r)+(u^r)^2}\equiv-4\pi N_0,\label{52}
\end{equation}
where $N_0=-N_1N_2$ and $N_2=(p_\infty+\rho\infty)\sqrt{f(r_\infty)}$ gives
\begin{equation}
\dot{M}=4\pi N_{1}(\rho_{\infty}+p{_\infty})\sqrt{f(r_{\infty})}M^2.\label{53}
\end{equation}
Here, we consider our boundary at infinity (i.e $r=r_{\infty}$).
The radius at infinity means special infinity where the massive particle falling into
the BH. Also, $\rho_{\infty}$ is $\rho$ (energy density) at $r_{\infty}$ and $p_{\infty}$ means $p$ (pressure) at $r_{\infty}$.
We take the time evolution of the mass of the BH,
for this, the above equation can be written in the following form
\begin{equation}
\frac{dM}{M^2}=  \mathcal{F}dt,\label{54}
\end{equation}
where $\mathcal{F}=4\pi N_{1}(\rho_{\infty}+p{_\infty})\sqrt{f(r_{\infty})}$, by integrating, we obtain
\begin{equation}
M_t=\frac{M_i}{1-Ft M_i}\equiv\frac{M_i}{1-\frac{t}{t_{cr}}},\label{55}
\end{equation}
where $M_i$ is the mass of BH with initial mass, $M_t$ is the mass of BH with the critical
accretion time and $t_{cr}=\left[4\pi N_{1}(\rho_{\infty}+p{_\infty})\sqrt{f(r_{\infty})M_i}\right]^{-1}$
is the critical accretion of time evolution. Therefore, the required expression of BH mass accretion rate is given by
\begin{equation}
\dot{M}=4\pi N_{1}(\rho+p)M^2.\label{56}
\end{equation}

\subsection{Critical accretion}
Since the fluid element is at rest far from the BH whereas it moves inwards, then it must be passed
through the critical point where the velocity of the moving fluid is equal to the sound speed.
The maximum accretion occurs if the moving fluid towards the critical point.
Taking $h=h(\rho)$ constant enthalpy then the fluid come to be barotropic.
For this, the equation of state is given by \cite{54}.
\begin{equation}
\frac{dh}{h} = V^2\frac{d\rho}{\rho},\label{51}
\end{equation}
where $V$ is the local speed of sound. Then this equation gives $\ln h=V^2\ln n$.
So, from Eqs. (\ref{44}), (\ref{45}) and (\ref{51}), we obtain
\begin{equation}
\left[\left(\frac{u}{u_t}\right)^2-V^2\right](\ln u)_{,r}=\frac{1}{r^2(u_t)}\left[2rV^2(u_t)^2-\frac{1}{2}r^2 f'(r)\right],\label{52}
\end{equation}
where critical points are denoted by the subscripted letter $c$ and
one can be found the solution of the local speed of sound at these points.
\begin{equation}
V^2_c = \left(\frac{u_c}{u_{tc}}\right)^2.\label{53}
\end{equation}
At the sonic points, we have
\begin{equation}
2rV^2_c(u_{tc})^2-\frac{1}{2}r^2_c f'_{rc} = 0,\label{54}
\end{equation}
where $f_c=f(r)|_{r=r_c}$ and $f'_{rc}=f'(r_c)$.
We obtain the radial velocity at the critical point by putting (\ref{53}) into (\ref{54}), given by
\begin{equation}
(u_c)^2 = \frac{1}{4r}r^2_c f'_{rc}.\label{55}
\end{equation}
By using Eqs. (\ref{34}), (\ref{54}) and (\ref{55}), we obtain
\begin{equation}
r^2_c f'_{rc} = 4rV^2_c[f(r_c)+(u_c)^2],\label{56}
\end{equation}
finally, it produces the local speed of sound, given by
\begin{equation}
V^2_c = \frac{r^2_c f'_{rc}}{r^2_c f'_{r_c}+4rf(r_c)}.\label{57}
\end{equation}

\section{Circular equatorial geodesics}
The explicit form of the effective potential is important for the proceeding of circular geodesics.
Hence, it is directed by Eq. (\ref{13}), give as
\begin{eqnarray}\label{58}
V^s_{eff}&=&\left(1-\frac{2M}{r}+\frac{Q^2}{r^2}-\frac{C^2k^2}{2r^2}+\frac{16C^{3/2}k^2}{15\beta^{1/4}r}\right)\left(1+\frac{L^2}{r^2}\right),\\
\label{59}
V^w_{eff}&=&\left(1-\frac{2M}{r}+\frac{Q^2}{r^2}-\frac{\beta C^4k^2}{10r^6}\right)\left(1+\frac{L^2}{r^2}\right),
\end{eqnarray}
where $\frac{d^2}{dr^2}V_{eff}>0$ is the condition for the presence of
the $ISCO$ and also the Eq. (\ref{20}) locate the
$ISCO$ at $r\geq3\left(M-\frac{8C^{3/2}k^2}{15\beta^{1/4}}+\sqrt{9\left(M-\frac{8C^{3/2}k^2}{15\beta^{1/4}}\right)^2-8Q^2+4C^2k^2}\right)$, as
\begin{equation}
r_{isco}=3\left(M-\frac{8C^{3/2}k^2}{15\beta^{1/4}}+\sqrt{9\left(M-\frac{8C^{3/2}k^2}{15\beta^{1/4}}\right)^2-8Q^2+4C^2k^2}\right),\label{60}
\end{equation}
it is the required characteristic radius of the $ISCO$ in the equatorial plane.
It has been noted that the $ISCO$ is an important study for the accretion process around the BH,
Also, some other circular orbits are necessary for the completion of this process.
Generally, the circular orbits proceed only when its radius is greater than the photon radius $r_{ph}$. For $r_{ph}<r<r_{mb}$,
the motion of the particle will be unstable means that the particle falls into the BH or flee away to infinity.
But in the region $r>r_{mb}$, the particle moves on the stable circular orbits.
The results of circular orbits such as photon sphere $r_{ph}$, circular orbit $r_{circ}$ and marginally
bound orbit $r_{mb}$ along with singularity $r_{sing}$ at $f(r)=0$ are given by
\begin{eqnarray}\label{61}
r_{sing}&=&\frac{1}{60}\left(60M-\frac{32C^{3/2}k^2}{\beta^{1/4}}+\frac{\sqrt{(32C^{3/2}k^2-60M\beta^{1/4})^2-120(-15C^2k^2\beta^{1/4}+30Q^2\beta^{1/4})
\beta^{1/4}}}{\beta^{1/4}}\right),\\
\label{62}
r_{ph}&=&\frac{3}{2}\left(M-\frac{8C^{3/2}k^2}{15\beta^{1/4}}+\frac{1}{2}\sqrt{9\left(M-\frac{8C^{3/2}k^2}{15\beta^{1/4}}\right)^2-8Q^2+4C^2k^2}\right),\\
\label{63}
r_{circ}&>&\frac{3}{2}\left(M-\frac{8C^{3/2}k^2}{15\beta^{1/4}}+\frac{1}{2}\sqrt{9\left(M-\frac{8C^{3/2}k^2}{15\beta^{1/4}}\right)^2-8Q^2+4C^2k^2}\right),\\
\label{64}
r_{mb}& =&\frac{1}{60}\left(60M-\frac{32C^{3/2}k^2}{\beta^{1/4}}+\frac{\sqrt{(-32C^{3/2}k^2+6015M\beta^{1/4})^2+120(15C^2k^2\beta^{1/4}-30Q^2\beta^{1/4})
\beta^{1/4}}}{\beta^{1/4}}\right).
\end{eqnarray}
Now, we calculate the specific energy, specific angular momentum, angular velocity, and angular momentum
of a moving particle in circular orbits for strong and weak fields. Therefore,
\begin{eqnarray}\label{64}
E_s^2&=&\frac{2\left(1-\frac{2M}{r}+\frac{Q^2}{r^2}-\frac{C^2k^2}{2r^2}+\frac{16C^{3/2}k^2}{15\beta^{1/4}r}\right)^2}
{2+\frac{2\left(-5C^2k^2+10Q^2-15Mr+\frac{8C^{3/2}k^2r}{\beta^{1/4}}\right)}{5r^2}}.\\
\label{65}
E_w^2&=&\frac{2\left(1-\frac{2M}{r}+\frac{Q^2}{r^2}-\frac{\beta C^4k^2}{10r^6}\right)^2}
{2+\frac{4Q^2}{r^2}-\frac{6M}{r}-\frac{4C^4k^2\beta}{5r^6}}.
\end{eqnarray}
\begin{eqnarray}\label{66}
L_s^2&=&\frac{C^2k^2-2Q^2+2Mr-\frac{16C^{3/2}k^2r}{15\beta^{1/4}}}
{2+\frac{2\left(-5C^2k^2+10Q^2-15Mr+\frac{8C^{3/2}k^2r}{\beta^{1/4}}\right)}{5r^2}}.\\
\label{67}
L_w^2&=&\frac{-2Q^2+2Mr+\frac{3C^4k^2\beta}{5r^4}}
{2+\frac{4Q^2}{r^2}-\frac{6M}{r}-\frac{4C^4k^2\beta}{5r^6}}.
\end{eqnarray}
\begin{eqnarray}\label{68}
\Omega^2_{\phi s} &=&\frac{1}{4r^2}\left[\frac{C^2k^2}{r^3}-\frac{2Q^2}{r^3}+\frac{2M}{r^2}-\frac{16C^{3/2}k^2}{15\beta^{1/4}r^2}\right]^2\\
\label{69}
\Omega^2_{\phi w} &=&\frac{1}{4r^2}\left[\frac{-2Q^2}{r^3}+\frac{2M}{r^2}+\frac{3C^4k^2\beta}{5r^7}\right]^2
\end{eqnarray}
\begin{eqnarray}\label{70}
l_s^2&=&\frac{C^2k^2-2Q^2+2Mr-\frac{16C^{3/2}k^2r}{15\beta^{1/4}}}
{2\left(1-\frac{2M}{r}+\frac{Q^2}{r^2}-\frac{C^2k^2}{2r^2}+\frac{16C^{3/2}k^2}{15\beta^{1/4}r}\right)^2}.\\
\label{71}
l_w^2&=&\frac{-2Q^2+2Mr+\frac{3C^4k^2\beta}{5r^4}}
{2\left(1-\frac{2M}{r}+\frac{Q^2}{r^2}-\frac{\beta C^4k^2}{10r^6}\right)^2}.
\end{eqnarray}

\subsection{Epicyclic frequencies}
If a particle is moving in a circular orbit, then it achieves small oscillations in the direction of radial
and vertical frequencies. These oscillations are the effects of perturbation on a moving particle in a circular orbit.
So, the required frequencies are given by
\begin{eqnarray}\label{72}
\Omega^2_{\theta s} &=&\frac{1}{r}\left[\frac{C^2k^2}{r^3}-\frac{2Q^2}{r^3}+\frac{2M}{r^2}-\frac{16C^{3/2}k^2}{15\beta^{1/4}r^2}\right]\\
\label{73}
\Omega^2_{\theta w} &=&\frac{1}{r}\left[\frac{-2Q^2}{r^3}+\frac{2M}{r^2}+\frac{3C^4k^2\beta}{5r^7}\right]
\end{eqnarray}
\begin{figure}
\centering \epsfig{file=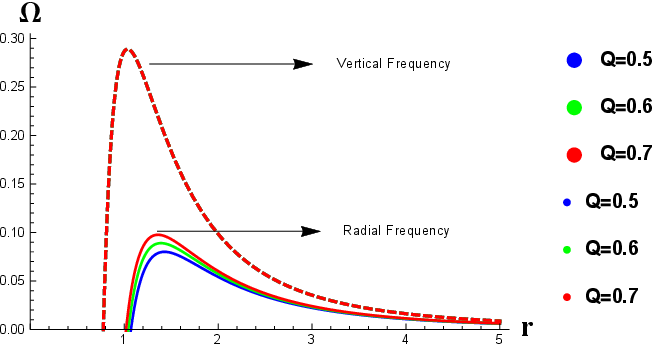, width=.45\linewidth,
height=2.2in}\epsfig{file=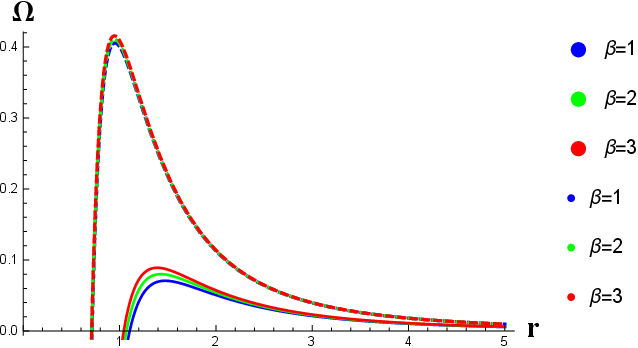, width=.45\linewidth,
height=2.2in}
\centering \epsfig{file=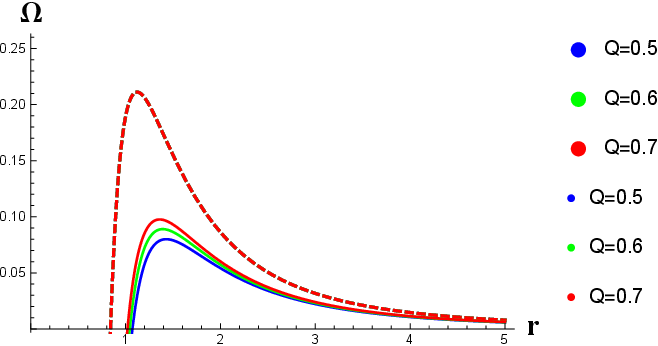, width=.45\linewidth,
height=2.02in}\epsfig{file=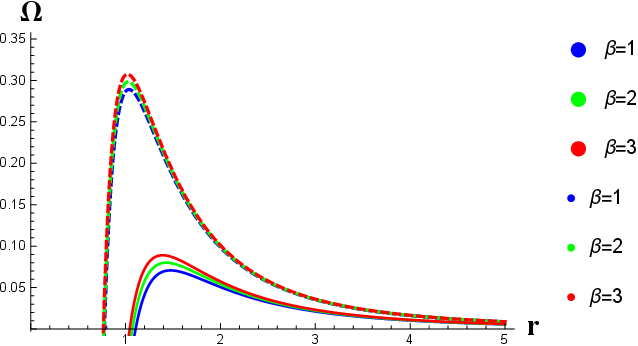, width=.45\linewidth,
height=2.02in}\caption{\label{F12} A variation between vertical and radial frequencies of strong and weak fields for $Q$ and $\beta$. Upper panels show the variation of frequency for the strong field while lower panels for the weak field.
The dependency of vertical frequency is on the thick dot values while the radial frequency is on the other dot values.
This behavior is for the equations (\ref{31}), (\ref{72}) and (\ref{73}).}
\end{figure}
The radial and vertical frequency variation is represented in Fig. (\ref{F12}) and has the following key points:
\begin{itemize}
  \item The vertical frequencies are shown in all panel of Fig. (\ref{F12}) by the dashed lines which are coincide and will decrease by increasing $r$.We see explicitly that this frequency is not dependent on parameters $Q$ and $\beta$.
  \item The radial epicyclic frequencies are shown by the solid curves with a maximum where increasing
        the parameters and shifts this maximum to the smaller radii $r=1$.
  \item Dependency on these parameters is significant near to the BH, but far from the BH this effect is weak.
  \item Clearly, we see that $\Omega_r<\Omega_\theta$.
\end{itemize}
\begin{figure}
\centering \epsfig{file=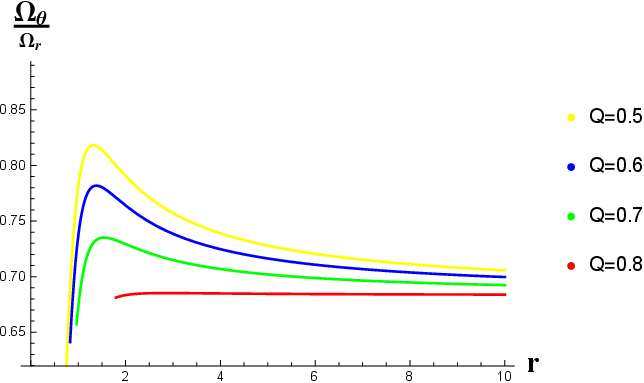, width=.45\linewidth,
height=2.2in}\epsfig{file=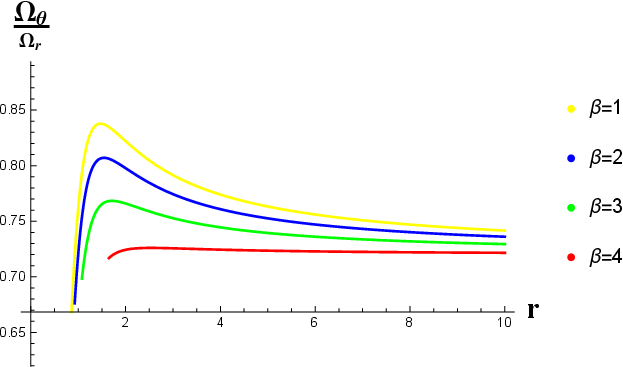, width=.45\linewidth,
height=2.2in}
\centering \epsfig{file=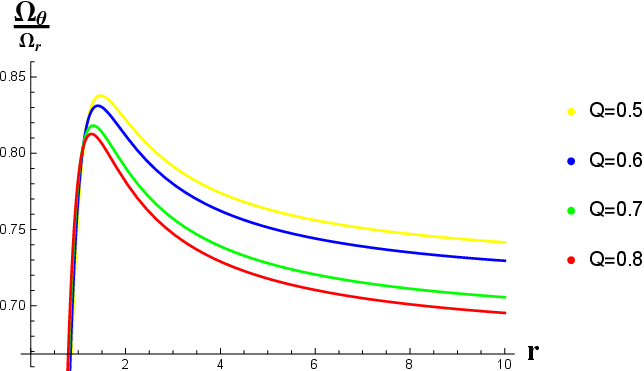, width=.45\linewidth,
height=2.02in}\epsfig{file=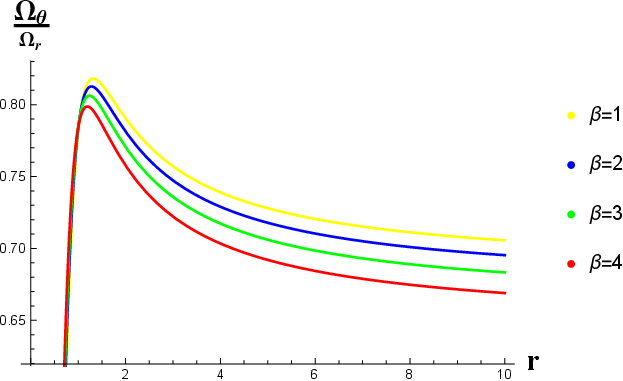, width=.45\linewidth,
height=2.02in}\caption{\label{F12} A variation between ratios of vertical and radial frequencies of strong and weak fields for $Q$ and $\beta$. Upper panels show the variation of frequency for the strong field while lower panels for the weak field.}
\end{figure}

The ratio $\frac{\Omega_\theta}{\Omega_r}$ in Fig. $13$, is a
decreasing function of $r$ . we noted that in the vicinity of the
BH, this ratio is very close to unity but far from
the BH it greater than unity and decreases by increasing the parameters $Q$ and $\beta$.

\section{Conclusion}
In this paper, we studied the geodesic motion and accretion process of a test particle's
near a nonlinear electrodynamic BH in strong and weak field approximation.
In this framework, we considered the equatorial plane and analyzed the circular geodesics with their stabilities,
oscillations for small perturbations, unstable orbits and
accretion of the fluid flowing onto the BH in a general form.
Further, the effective potential, specific energy, angular momentum,
epicyclic frequencies, characteristic radii, emission rate, and the mass evolution of the
BHs have been studied. Then some general solutions under the strong and weak field limits
are obtained by considering the equation of state $p=k\rho$ in the isothermal fluid.
\begin{itemize}
  \item The metric parameters of strong and weak fields suggest that the nonlinear electrodynamic effects
  can not be removed in the horizon structure but we can remove these effects for a far-distant observer.
The effects of parameters $Q$ and $\beta$ are considered for each case of strong and weak fields and some solutions
are compared to the Schwarzschild solution. The weak field analysis has shown that these solutions have a deviation
from the Schwarzschild solution (where recovered by $\beta=0$ and $Q=0$). In Fig. (\ref{F1}), Al the horizons
occur between the distance $r=(0,2)$ near the singularity, while for the maximum radius the curves are away from
the singularity. All the plots have two horizons except for the bottom right plot.

  \item In Fig. (\ref{F2}), the nonlinear electrodynamic parameters affect the effective potential, it can be seen
  that it is maximum for increased values of the parameters in a strong field. In a weak field,
  it maximizes for decreased values of the parameters, and the stable circular orbits are
   located at large distances while unstable circular orbits are located at small radii.
   We have noted the location of the characteristic radii ($r_{sing}, r_{ph}, r_{mb}, r_{isco}$) in
   Fig. (\ref{F3}). These radii have considerable deviation around the strong and weak fields.
   The unstable circular orbits fall onto the central mass at the distance $\beta=0.05$
   whereas the stable circular orbits are away from the central mass at the maximum $\beta$ in a strong field.
   The pictures of the radii clearly show that the radius $r_{ISCO}$ is greater than the other radii whereas
   the radius $r_{sing}$ is smaller than the other radii in strong and weak fields.
   The energy raises for increasing $Q$ and decreasing $\beta$ in a strong field. It can be seen from the energy diagram
  a very small deviation occur in the radius of the bound orbit. The angular momentum increases for decreasing $Q$
  and it raises for decreasing $\beta$ whereas the unstable orbits inward to $ISCO$ and stable orbits are outward to $ISCO$ in a strong field.
  In a weak field, the angular momentum raises for decreasing $Q$ and $\beta$ whereas the unstable orbits inward to $ISCO$ and stable orbits outward to $ISCO$.

  \item Increasing the parameters $Q$ and $\beta$, the efficiency of accretion decreases in a very small range ($0.04, 0.08$)
  in all plots for strong and weak fields. One can be seen that the unstable orbits have the maximum distances whereas the stable orbits
  have the minimum distances from the singularity. The epicyclic frequencies diagram shows that the radial frequency is greater than the
  vertical frequencies for increasing values of $Q$ and $\beta$ in strong and weak fields.
  It has been seen that the necessity of these variables is close
  to the central mass but extreme from the BH it is weak. From the behavior, see that $\Omega_r<\Omega_\theta$.
  The energy profile shows that it is increased for increasing the values of charge $Q$ and vice versa in a strong field.
  Also, it is decreased for increasing the values of $\beta$ and vice versa.
  In a weak field, the specific energy decreases, and a very small change in the bound orbit radius
  is noted for smaller values of $\beta$ and larger values of $Q$.

 \item The radiation flux decreases at the singularity and has
 a maximum position away from the singularity in the vicinity of the strong field.
 By increasing the charge $Q$, the flux decreases
 and the minimum of the flux turns to the singularity but
 the reverse happens for the parameter $\beta$.
 Both the parameters $Q$ and $\beta$ show the same pictures
 in the case of a weak field.
 Therefore, the dependence on these
 parameters is very important in the vicinity of the
 strong and weak fields. These behaviors
 are different for temperature. All the solution curves
 are bounded around the radius. The starting and ending points of these curves
 are in the bound radius $r$. Therefore, the maximum temperature
 happens for the value $Q=0$, and is decreased for other values.
\end{itemize}

In addition, we have investigated the radial velocity, energy density, and mass accretion rate
of strong and weak fields by considering isothermal fluid and equation of state parameter $\omega=1/2$.
It has been noted that the radial velocity raises at the smaller radii for both parameters in the vicinity of the strong field but
far from the field, the fluids have no radial velocity. Accretion takes place if the speed of fluids
equal to the speed of sound and the flow is subsonic before the critical point.
This flow will be supersonic around the BH so the speed of flow increases and
passes through the critical point in the locality of strong and weak fields.
The picture of a weak field represents that the radial velocity of the Schwarzschild BH
is greater than the nonlinear electrodynamics BH.
Therefore, the velocity increases by increasing the parameters,
the speed of flow equals the speed of sound nearby the BH.
From the density picture, it has been seen that the density of the fluid increases at larger radii
for both parameters in the vicinity of strong and weak fields. All the solution curves turn down
to the singularity and maximum density happens near to the singularity.
Finally, 
 analyzed the mass accretion rate, decreasing the parameters,
The accretion rate increases and it achieves the maximum position at $Q=0$ in the locality of a strong field.
All the solution curves turn down at larger radii and are bound in the right plot of a strong field.
In the locality of a weak field, the solution curves pass through the circular disk away from the singularity,
accretion rate increases for increasing the parameters.
Hence, the accretion rate depends on the metric parameters
and the nature of the fluid.

\bibliographystyle{apsrev4-1}  
\bibliography{Accretion}

\begin{thebibliography}{53}%
\makeatletter
\providecommand \@ifxundefined [1]{%
 \@ifx{#1\undefined}
}%
\providecommand \@ifnum [1]{%
 \ifnum #1\expandafter \@firstoftwo
 \else \expandafter \@secondoftwo
 \fi
}%
\providecommand \@ifx [1]{%
 \ifx #1\expandafter \@firstoftwo
 \else \expandafter \@secondoftwo
 \fi
}%
\providecommand \natexlab [1]{#1}%
\providecommand \enquote  [1]{``#1''}%
\providecommand \bibnamefont  [1]{#1}%
\providecommand \bibfnamefont [1]{#1}%
\providecommand \citenamefont [1]{#1}%
\providecommand \href@noop [0]{\@secondoftwo}%
\providecommand \href [0]{\begingroup \@sanitize@url \@href}%
\providecommand \@href[1]{\@@startlink{#1}\@@href}%
\providecommand \@@href[1]{\endgroup#1\@@endlink}%
\providecommand \@sanitize@url [0]{\catcode `\\12\catcode `\$12\catcode
  `\&12\catcode `\#12\catcode `\^12\catcode `\_12\catcode `\%12\relax}%
\providecommand \@@startlink[1]{}%
\providecommand \@@endlink[0]{}%
\providecommand \url  [0]{\begingroup\@sanitize@url \@url }%
\providecommand \@url [1]{\endgroup\@href {#1}{\urlprefix }}%
\providecommand \urlprefix  [0]{URL }%
\providecommand \Eprint [0]{\href }%
\providecommand \doibase [0]{http://dx.doi.org/}%
\providecommand \selectlanguage [0]{\@gobble}%
\providecommand \bibinfo  [0]{\@secondoftwo}%
\providecommand \bibfield  [0]{\@secondoftwo}%
\providecommand \translation [1]{[#1]}%
\providecommand \BibitemOpen [0]{}%
\providecommand \bibitemStop [0]{}%
\providecommand \bibitemNoStop [0]{.\EOS\space}%
\providecommand \EOS [0]{\spacefactor3000\relax}%
\providecommand \BibitemShut  [1]{\csname bibitem#1\endcsname}%
\let\auto@bib@innerbib\@empty
\bibitem [{\citenamefont {Abbott}\ and\ \citenamefont {et~al.}(2016)}]{1}%
  \BibitemOpen
  \bibfield  {author} {\bibinfo {author} {\bibfnamefont {B.~P.}\ \bibnamefont
  {Abbott}}\ and\ \bibinfo {author} {\bibnamefont {et~al.}} (\bibinfo
  {collaboration} {LIGO Scientific Collaboration and Virgo Collaboration}),\
  }\href {\doibase 10.1103/PhysRevLett.116.061102} {\bibfield  {journal}
  {\bibinfo  {journal} {Phys. Rev. Lett.}\ }\textbf {\bibinfo {volume} {116}},\
  \bibinfo {pages} {061102} (\bibinfo {year} {2016})}\BibitemShut {NoStop}%
\bibitem [{\citenamefont {{Akiyama}}\ and\ \citenamefont {et~al. {(Event
  Horizon Telescope Collaboration)}}(2019{\natexlab{a}})}]{2}%
  \BibitemOpen
  \bibfield  {author} {\bibinfo {author} {\bibfnamefont {K.}~\bibnamefont
  {{Akiyama}}}\ and\ \bibinfo {author} {\bibnamefont {et~al. {(Event Horizon
  Telescope Collaboration)}}},\ }\href {\doibase 10.3847/2041-8213/ab0ec7}
  {\bibfield  {journal} {\bibinfo  {journal} {Astrophys. J.}\ }\textbf
  {\bibinfo {volume} {875}},\ \bibinfo {eid} {L1} (\bibinfo {year}
  {2019}{\natexlab{a}})},\ \Eprint {http://arxiv.org/abs/1906.11238}
  {arXiv:1906.11238 [astro-ph.GA]} \BibitemShut {NoStop}%
\bibitem [{\citenamefont {{Akiyama}}\ and\ \citenamefont {et~al. {(Event
  Horizon Telescope Collaboration)}}(2019{\natexlab{b}})}]{3}%
  \BibitemOpen
  \bibfield  {author} {\bibinfo {author} {\bibfnamefont {K.}~\bibnamefont
  {{Akiyama}}}\ and\ \bibinfo {author} {\bibnamefont {et~al. {(Event Horizon
  Telescope Collaboration)}}},\ }\href {\doibase 10.3847/2041-8213/ab0e85}
  {\bibfield  {journal} {\bibinfo  {journal} {Astrophys. J.}\ }\textbf
  {\bibinfo {volume} {875}},\ \bibinfo {eid} {L4} (\bibinfo {year}
  {2019}{\natexlab{b}})},\ \Eprint {http://arxiv.org/abs/1906.11241}
  {arXiv:1906.11241 [astro-ph.GA]} \BibitemShut {NoStop}%
\bibitem [{\citenamefont {{Akiyama}}\ and\ \citenamefont {et~al. {(Event
  Horizon Telescope Collaboration)}}(2021)}]{4}%
  \BibitemOpen
  \bibfield  {author} {\bibinfo {author} {\bibfnamefont {K.}~\bibnamefont
  {{Akiyama}}}\ and\ \bibinfo {author} {\bibnamefont {et~al. {(Event Horizon
  Telescope Collaboration)}}},\ }\href {\doibase 10.3847/2041-8213/abe71d}
  {\bibfield  {journal} {\bibinfo  {journal} {Astrophys. J.}\ }\textbf
  {\bibinfo {volume} {910}},\ \bibinfo {eid} {L12} (\bibinfo {year}
  {2021})}\BibitemShut {NoStop}%
\bibitem [{\citenamefont {{Frank}}\ \emph {et~al.}(2002)\citenamefont
  {{Frank}}, \citenamefont {{King}},\ and\ \citenamefont {{Raine}}}]{5}%
  \BibitemOpen
  \bibfield  {author} {\bibinfo {author} {\bibfnamefont {J.}~\bibnamefont
  {{Frank}}}, \bibinfo {author} {\bibfnamefont {A.}~\bibnamefont {{King}}}, \
  and\ \bibinfo {author} {\bibfnamefont {D.~J.}\ \bibnamefont {{Raine}}},\
  }\href@noop {} {\emph {\bibinfo {title} {{Accretion Power in Astrophysics:
  Third Edition}}}}\ (\bibinfo {year} {2002})\BibitemShut {NoStop}%
\bibitem [{\citenamefont {Virbhadra}\ and\ \citenamefont {Ellis}(2000)}]{6}%
  \BibitemOpen
  \bibfield  {author} {\bibinfo {author} {\bibfnamefont {K.~S.}\ \bibnamefont
  {Virbhadra}}\ and\ \bibinfo {author} {\bibfnamefont {G.~F.~R.}\ \bibnamefont
  {Ellis}},\ }\href {\doibase 10.1103/PhysRevD.62.084003} {\bibfield  {journal}
  {\bibinfo  {journal} {Phys.~Rev.~D.}\ }\textbf {\bibinfo {volume} {62}},\
  \bibinfo {pages} {084003} (\bibinfo {year} {2000})}\BibitemShut {NoStop}%
\bibitem [{\citenamefont {{Nampalliwar}}\ and\ \citenamefont
  {{Bambi}}(2018)}]{7}%
  \BibitemOpen
  \bibfield  {author} {\bibinfo {author} {\bibfnamefont {S.}~\bibnamefont
  {{Nampalliwar}}}\ and\ \bibinfo {author} {\bibfnamefont {C.}~\bibnamefont
  {{Bambi}}},\ }\href@noop {} {\bibfield  {journal} {\bibinfo  {journal} {arXiv
  e-prints}\ ,\ \bibinfo {eid} {arXiv:1810.07041}} (\bibinfo {year} {2018})},\
  \Eprint {http://arxiv.org/abs/1810.07041} {arXiv:1810.07041 [astro-ph.HE]}
  \BibitemShut {NoStop}%
\bibitem [{\citenamefont {{Xie}}\ and\ \citenamefont {{Yuan}}(2012)}]{8}%
  \BibitemOpen
  \bibfield  {author} {\bibinfo {author} {\bibfnamefont {F.-G.}\ \bibnamefont
  {{Xie}}}\ and\ \bibinfo {author} {\bibfnamefont {F.}~\bibnamefont {{Yuan}}},\
  }\href {\doibase 10.1111/j.1365-2966.2012.22030.x} {\bibfield  {journal}
  {\bibinfo  {journal} {MNRAS}\ }\textbf {\bibinfo {volume} {427}},\ \bibinfo
  {pages} {1580} (\bibinfo {year} {2012})},\ \Eprint
  {http://arxiv.org/abs/1207.3113} {arXiv:1207.3113 [astro-ph.HE]} \BibitemShut
  {NoStop}%
\bibitem [{\citenamefont {{Kaplan}}(1949)}]{9}%
  \BibitemOpen
  \bibfield  {author} {\bibinfo {author} {\bibfnamefont {S.~A.}\ \bibnamefont
  {{Kaplan}}},\ }\href@noop {} {\bibfield  {journal} {\bibinfo  {journal}
  {JETP}\ }\textbf {\bibinfo {volume} {19}},\ \bibinfo {pages} {951} (\bibinfo
  {year} {1949})}\BibitemShut {NoStop}%
\bibitem [{\citenamefont {{Kato}}\ \emph {et~al.}(2008)\citenamefont {{Kato}},
  \citenamefont {{Fukue}},\ and\ \citenamefont {{Mineshige}}}]{10}%
  \BibitemOpen
  \bibfield  {author} {\bibinfo {author} {\bibfnamefont {S.}~\bibnamefont
  {{Kato}}}, \bibinfo {author} {\bibfnamefont {J.}~\bibnamefont {{Fukue}}}, \
  and\ \bibinfo {author} {\bibfnamefont {S.}~\bibnamefont {{Mineshige}}},\
  }\href@noop {} {\emph {\bibinfo {title} {{Black-Hole Accretion Disks ---
  Towards a New Paradigm ---}}}}\ (\bibinfo {year} {2008})\BibitemShut
  {NoStop}%
\bibitem [{\citenamefont {{Kato}}(2001)}]{11}%
  \BibitemOpen
  \bibfield  {author} {\bibinfo {author} {\bibfnamefont {S.}~\bibnamefont
  {{Kato}}},\ }\href {\doibase 10.1093/pasj/53.5.L37} {\bibfield  {journal}
  {\bibinfo  {journal} {Astron. Soc. Jpn.}\ }\textbf {\bibinfo {volume} {53}},\
  \bibinfo {pages} {L37} (\bibinfo {year} {2001})}\BibitemShut {NoStop}%
\bibitem [{\citenamefont {{Ortega-Rodr{\'\i}guez}}\ \emph
  {et~al.}(2008)\citenamefont {{Ortega-Rodr{\'\i}guez}}, \citenamefont
  {{Silbergleit}},\ and\ \citenamefont {{Wagoner}}}]{12}%
  \BibitemOpen
  \bibfield  {author} {\bibinfo {author} {\bibfnamefont {M.}~\bibnamefont
  {{Ortega-Rodr{\'\i}guez}}}, \bibinfo {author} {\bibfnamefont
  {A.}~\bibnamefont {{Silbergleit}}}, \ and\ \bibinfo {author} {\bibfnamefont
  {R.}~\bibnamefont {{Wagoner}}},\ }\href {\doibase 10.1080/03091920701462130}
  {\bibfield  {journal} {\bibinfo  {journal} {Astrophys. Fluid Dyn.}\ }\textbf
  {\bibinfo {volume} {102}},\ \bibinfo {pages} {75} (\bibinfo {year} {2008})},\
  \Eprint {http://arxiv.org/abs/astro-ph/0611101} {arXiv:astro-ph/0611101
  [astro-ph]} \BibitemShut {NoStop}%
\bibitem [{\citenamefont {{Kluzniak}}\ and\ \citenamefont
  {{Abramowicz}}(2001)}]{13}%
  \BibitemOpen
  \bibfield  {author} {\bibinfo {author} {\bibfnamefont {W.}~\bibnamefont
  {{Kluzniak}}}\ and\ \bibinfo {author} {\bibfnamefont {M.~A.}\ \bibnamefont
  {{Abramowicz}}},\ }\href@noop {} {\bibfield  {journal} {\bibinfo  {journal}
  {arXiv e-prints}\ ,\ \bibinfo {eid} {astro-ph/0105057}} (\bibinfo {year}
  {2001})},\ \Eprint {http://arxiv.org/abs/astro-ph/0105057}
  {arXiv:astro-ph/0105057 [astro-ph]} \BibitemShut {NoStop}%
\bibitem [{\citenamefont {{Landau}}\ and\ \citenamefont
  {{Lifshitz}}(1975)}]{14}%
  \BibitemOpen
  \bibfield  {author} {\bibinfo {author} {\bibfnamefont {L.~D.}\ \bibnamefont
  {{Landau}}}\ and\ \bibinfo {author} {\bibfnamefont {E.~M.}\ \bibnamefont
  {{Lifshitz}}},\ }\href@noop {} {\emph {\bibinfo {title} {{The classical
  theory of fields}}}}\ (\bibinfo {year} {1975})\BibitemShut {NoStop}%
\bibitem [{\citenamefont {{Ruffini}}\ and\ \citenamefont
  {{Wheeler}}(1971)}]{15}%
  \BibitemOpen
  \bibfield  {author} {\bibinfo {author} {\bibfnamefont {R.}~\bibnamefont
  {{Ruffini}}}\ and\ \bibinfo {author} {\bibfnamefont {J.~A.}\ \bibnamefont
  {{Wheeler}}},\ }\href@noop {} {\bibfield  {journal} {\bibinfo  {journal}
  {ESRO}\ }\textbf {\bibinfo {volume} {52}},\ \bibinfo {pages} {45} (\bibinfo
  {year} {1971})}\BibitemShut {NoStop}%
\bibitem [{\citenamefont {{Bardeen}}\ \emph {et~al.}(1972)\citenamefont
  {{Bardeen}}, \citenamefont {{Press}},\ and\ \citenamefont
  {{Teukolsky}}}]{16}%
  \BibitemOpen
  \bibfield  {author} {\bibinfo {author} {\bibfnamefont {J.~M.}\ \bibnamefont
  {{Bardeen}}}, \bibinfo {author} {\bibfnamefont {W.~H.}\ \bibnamefont
  {{Press}}}, \ and\ \bibinfo {author} {\bibfnamefont {S.~A.}\ \bibnamefont
  {{Teukolsky}}},\ }\href {\doibase 10.1086/151796} {\bibfield  {journal}
  {\bibinfo  {journal} {Astrophys. J.}\ }\textbf {\bibinfo {volume} {178}},\
  \bibinfo {pages} {347} (\bibinfo {year} {1972})}\BibitemShut {NoStop}%
\bibitem [{\citenamefont {{Hobson}}\ \emph {et~al.}(2006)\citenamefont
  {{Hobson}}, \citenamefont {{Efstathiou}},\ and\ \citenamefont
  {{Lasenby}}}]{17}%
  \BibitemOpen
  \bibfield  {author} {\bibinfo {author} {\bibfnamefont {M.~P.}\ \bibnamefont
  {{Hobson}}}, \bibinfo {author} {\bibfnamefont {G.~P.}\ \bibnamefont
  {{Efstathiou}}}, \ and\ \bibinfo {author} {\bibfnamefont {A.~N.}\
  \bibnamefont {{Lasenby}}},\ }\href {\doibase 10.2277/0521829518} {\emph
  {\bibinfo {title} {{General Relativity}}}}\ (\bibinfo {year}
  {2006})\BibitemShut {NoStop}%
\bibitem [{\citenamefont {{Novikov}}\ and\ \citenamefont
  {{Thorne}}(1973)}]{18}%
  \BibitemOpen
  \bibfield  {author} {\bibinfo {author} {\bibfnamefont {I.~D.}\ \bibnamefont
  {{Novikov}}}\ and\ \bibinfo {author} {\bibfnamefont {K.~S.}\ \bibnamefont
  {{Thorne}}},\ }in\ \href@noop {} {\emph {\bibinfo {booktitle} {Black Holes
  (Les Astres Occlus)}}}\ (\bibinfo {year} {1973})\ pp.\ \bibinfo {pages}
  {343--450}\BibitemShut {NoStop}%
\bibitem [{\citenamefont {{Johannsen}}\ and\ \citenamefont
  {{Psaltis}}(2011)}]{19}%
  \BibitemOpen
  \bibfield  {author} {\bibinfo {author} {\bibfnamefont {T.}~\bibnamefont
  {{Johannsen}}}\ and\ \bibinfo {author} {\bibfnamefont {D.}~\bibnamefont
  {{Psaltis}}},\ }\href {\doibase 10.1103/PhysRevD.83.124015} {\bibfield
  {journal} {\bibinfo  {journal} {Phys. Rev. D}\ }\textbf {\bibinfo {volume}
  {83}},\ \bibinfo {eid} {124015} (\bibinfo {year} {2011})},\ \Eprint
  {http://arxiv.org/abs/1105.3191} {arXiv:1105.3191 [gr-qc]} \BibitemShut
  {NoStop}%
\bibitem [{\citenamefont {{Johannsen}}(2013)}]{20}%
  \BibitemOpen
  \bibfield  {author} {\bibinfo {author} {\bibfnamefont {T.}~\bibnamefont
  {{Johannsen}}},\ }\href {\doibase 10.1103/PhysRevD.87.124010} {\bibfield
  {journal} {\bibinfo  {journal} {Phys. Rev. D}\ }\textbf {\bibinfo {volume}
  {87}},\ \bibinfo {eid} {124010} (\bibinfo {year} {2013})},\ \Eprint
  {http://arxiv.org/abs/1304.8106} {arXiv:1304.8106 [gr-qc]} \BibitemShut
  {NoStop}%
\bibitem [{\citenamefont {{Tursunov}}\ \emph {et~al.}(2016)\citenamefont
  {{Tursunov}}, \citenamefont {{Stuchl{\'\i}k}},\ and\ \citenamefont
  {{Kolo{\v{s}}}}}]{21}%
  \BibitemOpen
  \bibfield  {author} {\bibinfo {author} {\bibfnamefont {A.}~\bibnamefont
  {{Tursunov}}}, \bibinfo {author} {\bibfnamefont {Z.}~\bibnamefont
  {{Stuchl{\'\i}k}}}, \ and\ \bibinfo {author} {\bibfnamefont {M.}~\bibnamefont
  {{Kolo{\v{s}}}}},\ }\href {\doibase 10.1103/PhysRevD.93.084012} {\bibfield
  {journal} {\bibinfo  {journal} {Phys. Rev. D}\ }\textbf {\bibinfo {volume}
  {93}},\ \bibinfo {eid} {084012} (\bibinfo {year} {2016})},\ \Eprint
  {http://arxiv.org/abs/1603.07264} {arXiv:1603.07264 [gr-qc]} \BibitemShut
  {NoStop}%
\bibitem [{\citenamefont {{Ipser}}(1996)}]{22}%
  \BibitemOpen
  \bibfield  {author} {\bibinfo {author} {\bibfnamefont {J.~R.}\ \bibnamefont
  {{Ipser}}},\ }\href {\doibase 10.1086/176832} {\bibfield  {journal} {\bibinfo
   {journal} {Astrophys. J.}\ }\textbf {\bibinfo {volume} {458}},\ \bibinfo
  {pages} {508} (\bibinfo {year} {1996})}\BibitemShut {NoStop}%
\bibitem [{\citenamefont {{Ipser}}(1994)}]{23}%
  \BibitemOpen
  \bibfield  {author} {\bibinfo {author} {\bibfnamefont {J.~R.}\ \bibnamefont
  {{Ipser}}},\ }\href {\doibase 10.1086/174854} {\bibfield  {journal} {\bibinfo
   {journal} {Astrophys. J.}\ }\textbf {\bibinfo {volume} {435}},\ \bibinfo
  {pages} {767} (\bibinfo {year} {1994})}\BibitemShut {NoStop}%
\bibitem [{\citenamefont {{Wagoner}}(1999)}]{24}%
  \BibitemOpen
  \bibfield  {author} {\bibinfo {author} {\bibfnamefont {R.~V.}\ \bibnamefont
  {{Wagoner}}},\ }\href {\doibase 10.1016/S0370-1573(98)00104-5} {\bibfield
  {journal} {\bibinfo  {journal} {Phys. Rev.}\ }\textbf {\bibinfo {volume}
  {311}},\ \bibinfo {pages} {259} (\bibinfo {year} {1999})},\ \Eprint
  {http://arxiv.org/abs/astro-ph/9805028} {arXiv:astro-ph/9805028 [astro-ph]}
  \BibitemShut {NoStop}%
\bibitem [{\citenamefont {{Bondi}}(1952)}]{25}%
  \BibitemOpen
  \bibfield  {author} {\bibinfo {author} {\bibfnamefont {H.}~\bibnamefont
  {{Bondi}}},\ }\href {\doibase 10.1093/mnras/112.2.195} {\bibfield  {journal}
  {\bibinfo  {journal} {Mon. Not. R. Astron. Soc.}\ }\textbf {\bibinfo {volume}
  {112}},\ \bibinfo {pages} {195} (\bibinfo {year} {1952})}\BibitemShut
  {NoStop}%
\bibitem [{\citenamefont {{Rezzolla}}\ and\ \citenamefont
  {{Zanotti}}(2013)}]{26}%
  \BibitemOpen
  \bibfield  {author} {\bibinfo {author} {\bibfnamefont {L.}~\bibnamefont
  {{Rezzolla}}}\ and\ \bibinfo {author} {\bibfnamefont {O.}~\bibnamefont
  {{Zanotti}}},\ }\href@noop {} {\emph {\bibinfo {title} {{Relativistic
  Hydrodynamics}}}}\ (\bibinfo {year} {2013})\BibitemShut {NoStop}%
\bibitem [{\citenamefont {{Michel}}(1972)}]{27}%
  \BibitemOpen
  \bibfield  {author} {\bibinfo {author} {\bibfnamefont {F.~C.}\ \bibnamefont
  {{Michel}}},\ }\href {\doibase 10.1007/BF00649949} {\bibfield  {journal}
  {\bibinfo  {journal} {Astrophys. Space Sci.}\ }\textbf {\bibinfo {volume}
  {15}},\ \bibinfo {pages} {153} (\bibinfo {year} {1972})}\BibitemShut
  {NoStop}%
\bibitem [{\citenamefont {{Bondi}}\ and\ \citenamefont {{Hoyle}}(1944)}]{28}%
  \BibitemOpen
  \bibfield  {author} {\bibinfo {author} {\bibfnamefont {H.}~\bibnamefont
  {{Bondi}}}\ and\ \bibinfo {author} {\bibfnamefont {F.}~\bibnamefont
  {{Hoyle}}},\ }\href {\doibase 10.1093/mnras/104.5.273} {\bibfield  {journal}
  {\bibinfo  {journal} {Mon. Not. R. Astron. Soc.}\ }\textbf {\bibinfo {volume}
  {104}},\ \bibinfo {pages} {273} (\bibinfo {year} {1944})}\BibitemShut
  {NoStop}%
\bibitem [{\citenamefont {{Hoyle}}\ and\ \citenamefont
  {{Lyttleton}}(1939)}]{29}%
  \BibitemOpen
  \bibfield  {author} {\bibinfo {author} {\bibfnamefont {F.}~\bibnamefont
  {{Hoyle}}}\ and\ \bibinfo {author} {\bibfnamefont {R.~A.}\ \bibnamefont
  {{Lyttleton}}},\ }\href {\doibase 10.1017/S0305004100021150} {\bibfield
  {journal} {\bibinfo  {journal} {Proceedings of the Cambridge Philosophical
  Society}\ }\textbf {\bibinfo {volume} {35}},\ \bibinfo {pages} {405}
  (\bibinfo {year} {1939})}\BibitemShut {NoStop}%
\bibitem [{\citenamefont {{Tejeda}}\ and\ \citenamefont
  {{Aguayo-Ortiz}}(2019)}]{30}%
  \BibitemOpen
  \bibfield  {author} {\bibinfo {author} {\bibfnamefont {E.}~\bibnamefont
  {{Tejeda}}}\ and\ \bibinfo {author} {\bibfnamefont {A.}~\bibnamefont
  {{Aguayo-Ortiz}}},\ }\href {\doibase 10.1093/mnras/stz1513} {\bibfield
  {journal} {\bibinfo  {journal} {Mon. Not. R. Astron. Soc.}\ }\textbf
  {\bibinfo {volume} {487}},\ \bibinfo {pages} {3607} (\bibinfo {year}
  {2019})},\ \Eprint {http://arxiv.org/abs/1906.04923} {arXiv:1906.04923
  [astro-ph.HE]} \BibitemShut {NoStop}%
\bibitem [{\citenamefont {{Karkowski}}\ and\ \citenamefont
  {{Malec}}(2013)}]{31}%
  \BibitemOpen
  \bibfield  {author} {\bibinfo {author} {\bibfnamefont {J.}~\bibnamefont
  {{Karkowski}}}\ and\ \bibinfo {author} {\bibfnamefont {E.}~\bibnamefont
  {{Malec}}},\ }\href {\doibase 10.1103/PhysRevD.87.044007} {\bibfield
  {journal} {\bibinfo  {journal} {Phys. Rev. D}\ }\textbf {\bibinfo {volume}
  {87}},\ \bibinfo {eid} {044007} (\bibinfo {year} {2013})},\ \Eprint
  {http://arxiv.org/abs/1211.3618} {arXiv:1211.3618 [gr-qc]} \BibitemShut
  {NoStop}%
\bibitem [{\citenamefont {{Mach}}\ and\ \citenamefont {{Malec}}(2013)}]{32}%
  \BibitemOpen
  \bibfield  {author} {\bibinfo {author} {\bibfnamefont {P.}~\bibnamefont
  {{Mach}}}\ and\ \bibinfo {author} {\bibfnamefont {E.}~\bibnamefont
  {{Malec}}},\ }\href {\doibase 10.1103/PhysRevD.88.084055} {\bibfield
  {journal} {\bibinfo  {journal} {Phys. Rev. D}\ }\textbf {\bibinfo {volume}
  {88}},\ \bibinfo {eid} {084055} (\bibinfo {year} {2013})},\ \Eprint
  {http://arxiv.org/abs/1309.1546} {arXiv:1309.1546 [gr-qc]} \BibitemShut
  {NoStop}%
\bibitem [{\citenamefont {{Chaverra}}\ and\ \citenamefont
  {{Sarbach}}(2015)}]{33}%
  \BibitemOpen
  \bibfield  {author} {\bibinfo {author} {\bibfnamefont {E.}~\bibnamefont
  {{Chaverra}}}\ and\ \bibinfo {author} {\bibfnamefont {O.}~\bibnamefont
  {{Sarbach}}},\ }\href {\doibase 10.1088/0264-9381/32/15/155006} {\bibfield
  {journal} {\bibinfo  {journal} {Classical and Quantum Gravity}\ }\textbf
  {\bibinfo {volume} {32}},\ \bibinfo {eid} {155006} (\bibinfo {year}
  {2015})},\ \Eprint {http://arxiv.org/abs/1501.01641} {arXiv:1501.01641
  [gr-qc]} \BibitemShut {NoStop}%
\bibitem [{\citenamefont {{Hunt}}(1971)}]{34}%
  \BibitemOpen
  \bibfield  {author} {\bibinfo {author} {\bibfnamefont {R.}~\bibnamefont
  {{Hunt}}},\ }\href {\doibase 10.1093/mnras/154.2.141} {\bibfield  {journal}
  {\bibinfo  {journal} {Mon. Not. R. Astron. Soc.}\ }\textbf {\bibinfo {volume}
  {154}},\ \bibinfo {pages} {141} (\bibinfo {year} {1971})}\BibitemShut
  {NoStop}%
\bibitem [{\citenamefont {{Ruffert}}(1994)}]{35}%
  \BibitemOpen
  \bibfield  {author} {\bibinfo {author} {\bibfnamefont {M.}~\bibnamefont
  {{Ruffert}}},\ }\href {\doibase 10.1086/174144} {\bibfield  {journal}
  {\bibinfo  {journal} {Astrophys. J.}\ }\textbf {\bibinfo {volume} {427}},\
  \bibinfo {pages} {342} (\bibinfo {year} {1994})}\BibitemShut {NoStop}%
\bibitem [{\citenamefont {{Lora-Clavijo}}\ and\ \citenamefont
  {{Guzm{\'a}n}}(2013)}]{36}%
  \BibitemOpen
  \bibfield  {author} {\bibinfo {author} {\bibfnamefont {F.~D.}\ \bibnamefont
  {{Lora-Clavijo}}}\ and\ \bibinfo {author} {\bibfnamefont {F.~S.}\
  \bibnamefont {{Guzm{\'a}n}}},\ }\href {\doibase 10.1093/mnras/sts573}
  {\bibfield  {journal} {\bibinfo  {journal} {Mon. Not. R. Astron. Soc.}\
  }\textbf {\bibinfo {volume} {429}},\ \bibinfo {pages} {3144} (\bibinfo {year}
  {2013})},\ \Eprint {http://arxiv.org/abs/1212.2139} {arXiv:1212.2139
  [astro-ph.HE]} \BibitemShut {NoStop}%
\bibitem [{\citenamefont {{Abbas}}\ and\ \citenamefont {{Ditta}}(2018)}]{37}%
  \BibitemOpen
  \bibfield  {author} {\bibinfo {author} {\bibfnamefont {G.}~\bibnamefont
  {{Abbas}}}\ and\ \bibinfo {author} {\bibfnamefont {A.}~\bibnamefont
  {{Ditta}}},\ }\href {\doibase 10.1142/S0217732318500700} {\bibfield
  {journal} {\bibinfo  {journal} {Mod. Phys. Lett. A}\ }\textbf {\bibinfo
  {volume} {33}},\ \bibinfo {eid} {1850070} (\bibinfo {year}
  {2018})}\BibitemShut {NoStop}%
\bibitem [{\citenamefont {{Abbas}}\ and\ \citenamefont {{Ditta}}(2019)}]{38}%
  \BibitemOpen
  \bibfield  {author} {\bibinfo {author} {\bibfnamefont {G.}~\bibnamefont
  {{Abbas}}}\ and\ \bibinfo {author} {\bibfnamefont {A.}~\bibnamefont
  {{Ditta}}},\ }\href {\doibase 10.1007/s10714-019-2527-0} {\bibfield
  {journal} {\bibinfo  {journal} {Gen. Relat. Grav.}\ }\textbf {\bibinfo
  {volume} {51}},\ \bibinfo {eid} {43} (\bibinfo {year} {2019})}\BibitemShut
  {NoStop}%
\bibitem [{\citenamefont {{Abbas}}\ \emph {et~al.}(2019)\citenamefont
  {{Abbas}}, \citenamefont {{Ditta}}, \citenamefont {{Jawad}},\ and\
  \citenamefont {{Umair Shahzad}}}]{39}%
  \BibitemOpen
  \bibfield  {author} {\bibinfo {author} {\bibfnamefont {G.}~\bibnamefont
  {{Abbas}}}, \bibinfo {author} {\bibfnamefont {A.}~\bibnamefont {{Ditta}}},
  \bibinfo {author} {\bibfnamefont {A.}~\bibnamefont {{Jawad}}}, \ and\
  \bibinfo {author} {\bibfnamefont {M.}~\bibnamefont {{Umair Shahzad}}},\
  }\href {\doibase 10.1007/s10714-019-2620-4} {\bibfield  {journal} {\bibinfo
  {journal} {Gen. Relat. Grav.}\ }\textbf {\bibinfo {volume} {51}},\ \bibinfo
  {eid} {136} (\bibinfo {year} {2019})}\BibitemShut {NoStop}%
\bibitem [{\citenamefont {{Abbas}}\ and\ \citenamefont {{Ditta}}(2020)}]{40}%
  \BibitemOpen
  \bibfield  {author} {\bibinfo {author} {\bibfnamefont {G.}~\bibnamefont
  {{Abbas}}}\ and\ \bibinfo {author} {\bibfnamefont {A.}~\bibnamefont
  {{Ditta}}},\ }\href {\doibase 10.1140/epjc/s10052-020-08787-x} {\bibfield
  {journal} {\bibinfo  {journal} {Eur. Phys. J. C}\ }\textbf {\bibinfo {volume}
  {80}},\ \bibinfo {eid} {1212} (\bibinfo {year} {2020})},\ \Eprint
  {http://arxiv.org/abs/2012.12035} {arXiv:2012.12035 [gr-qc]} \BibitemShut
  {NoStop}%
\bibitem [{\citenamefont {{Ditta}}\ and\ \citenamefont
  {{Abbas}}(2020{\natexlab{a}})}]{41}%
  \BibitemOpen
  \bibfield  {author} {\bibinfo {author} {\bibfnamefont {A.}~\bibnamefont
  {{Ditta}}}\ and\ \bibinfo {author} {\bibfnamefont {G.}~\bibnamefont
  {{Abbas}}},\ }\href {\doibase 10.1007/s10714-020-02724-9} {\bibfield
  {journal} {\bibinfo  {journal} {Gen. Relat. Grav.}\ }\textbf {\bibinfo
  {volume} {52}},\ \bibinfo {eid} {77} (\bibinfo {year}
  {2020}{\natexlab{a}})}\BibitemShut {NoStop}%
\bibitem [{\citenamefont {{Abbas}}\ \emph
  {et~al.}(2021{\natexlab{a}})\citenamefont {{Abbas}}, \citenamefont {{Azam}},\
  and\ \citenamefont {{Ditta}}}]{42}%
  \BibitemOpen
  \bibfield  {author} {\bibinfo {author} {\bibfnamefont {G.}~\bibnamefont
  {{Abbas}}}, \bibinfo {author} {\bibfnamefont {M.}~\bibnamefont {{Azam}}}, \
  and\ \bibinfo {author} {\bibfnamefont {A.}~\bibnamefont {{Ditta}}},\ }\href
  {\doibase 10.1016/j.cjph.2020.10.032} {\bibfield  {journal} {\bibinfo
  {journal} {Chinese Journal of Physics}\ }\textbf {\bibinfo {volume} {69}},\
  \bibinfo {pages} {143} (\bibinfo {year} {2021}{\natexlab{a}})}\BibitemShut
  {NoStop}%
\bibitem [{\citenamefont {{Ditta}}\ and\ \citenamefont
  {{Abbas}}(2020{\natexlab{b}})}]{43}%
  \BibitemOpen
  \bibfield  {author} {\bibinfo {author} {\bibfnamefont {A.}~\bibnamefont
  {{Ditta}}}\ and\ \bibinfo {author} {\bibfnamefont {G.}~\bibnamefont
  {{Abbas}}},\ }\href {\doibase 10.1016/j.newast.2020.101437} {\bibfield
  {journal} {\bibinfo  {journal} {New Astronomy}\ }\textbf {\bibinfo {volume}
  {81}},\ \bibinfo {eid} {101437} (\bibinfo {year}
  {2020}{\natexlab{b}})}\BibitemShut {NoStop}%
\bibitem [{\citenamefont {{Abbas}}\ \emph
  {et~al.}(2021{\natexlab{b}})\citenamefont {{Abbas}}, \citenamefont {{Azam}},\
  and\ \citenamefont {{Ditta}}}]{44}%
  \BibitemOpen
  \bibfield  {author} {\bibinfo {author} {\bibfnamefont {G.}~\bibnamefont
  {{Abbas}}}, \bibinfo {author} {\bibfnamefont {M.}~\bibnamefont {{Azam}}}, \
  and\ \bibinfo {author} {\bibfnamefont {A.}~\bibnamefont {{Ditta}}},\ }\href
  {\doibase 10.1016/j.cjph.2020.10.032} {\bibfield  {journal} {\bibinfo
  {journal} {Chinese Journal of Physics}\ }\textbf {\bibinfo {volume} {69}},\
  \bibinfo {pages} {143} (\bibinfo {year} {2021}{\natexlab{b}})}\BibitemShut
  {NoStop}%
\bibitem [{\citenamefont {{Abbas}}\ and\ \citenamefont {{Ditta}}(2021)}]{45}%
  \BibitemOpen
  \bibfield  {author} {\bibinfo {author} {\bibfnamefont {G.}~\bibnamefont
  {{Abbas}}}\ and\ \bibinfo {author} {\bibfnamefont {A.}~\bibnamefont
  {{Ditta}}},\ }\href {\doibase 10.1016/j.newast.2020.101508} {\bibfield
  {journal} {\bibinfo  {journal} {New Astronomy}\ }\textbf {\bibinfo {volume}
  {84}},\ \bibinfo {eid} {101508} (\bibinfo {year} {2021})}\BibitemShut
  {NoStop}%
\bibitem [{\citenamefont {{Tretyakova}}(2016)}]{46}%
  \BibitemOpen
  \bibfield  {author} {\bibinfo {author} {\bibfnamefont {D.~A.}\ \bibnamefont
  {{Tretyakova}}},\ }\href@noop {} {\bibfield  {journal} {\bibinfo  {journal}
  {arXiv e-prints}\ ,\ \bibinfo {eid} {arXiv:1606.08569}} (\bibinfo {year}
  {2016})},\ \Eprint {http://arxiv.org/abs/1606.08569} {arXiv:1606.08569
  [gr-qc]} \BibitemShut {NoStop}%
\bibitem [{\citenamefont {{Salahshoor}}\ and\ \citenamefont
  {{Nozari}}(2018)}]{47}%
  \BibitemOpen
  \bibfield  {author} {\bibinfo {author} {\bibfnamefont {K.}~\bibnamefont
  {{Salahshoor}}}\ and\ \bibinfo {author} {\bibfnamefont {K.}~\bibnamefont
  {{Nozari}}},\ }\href {\doibase 10.1140/epjc/s10052-018-5946-2} {\bibfield
  {journal} {\bibinfo  {journal} {European Physical Journal C}\ }\textbf
  {\bibinfo {volume} {78}},\ \bibinfo {eid} {486} (\bibinfo {year} {2018})},\
  \Eprint {http://arxiv.org/abs/1806.08949} {arXiv:1806.08949 [gr-qc]}
  \BibitemShut {NoStop}%
\bibitem [{\citenamefont {{Kruglov}}(2015)}]{48}%
  \BibitemOpen
  \bibfield  {author} {\bibinfo {author} {\bibfnamefont {S.~I.}\ \bibnamefont
  {{Kruglov}}},\ }\href {\doibase 10.1142/S0219887815500735} {\bibfield
  {journal} {\bibinfo  {journal} {Int. J. Geom. Meth. Mod. Phys.}\ }\textbf
  {\bibinfo {volume} {12}},\ \bibinfo {eid} {1550073} (\bibinfo {year}
  {2015})},\ \Eprint {http://arxiv.org/abs/1504.03941} {arXiv:1504.03941
  [physics.gen-ph]} \BibitemShut {NoStop}%
\bibitem [{\citenamefont {{Uniyal}}\ \emph {et~al.}(2022)\citenamefont
  {{Uniyal}}, \citenamefont {{Pantig}},\ and\ \citenamefont
  {{{\"O}vg{\"u}n}}}]{49}%
  \BibitemOpen
  \bibfield  {author} {\bibinfo {author} {\bibfnamefont {A.}~\bibnamefont
  {{Uniyal}}}, \bibinfo {author} {\bibfnamefont {R.~C.}\ \bibnamefont
  {{Pantig}}}, \ and\ \bibinfo {author} {\bibfnamefont {A.}~\bibnamefont
  {{{\"O}vg{\"u}n}}},\ }\href@noop {} {\bibfield  {journal} {\bibinfo
  {journal} {arXiv e-prints}\ ,\ \bibinfo {eid} {arXiv:2205.11072}} (\bibinfo
  {year} {2022})},\ \Eprint {http://arxiv.org/abs/2205.11072} {arXiv:2205.11072
  [gr-qc]} \BibitemShut {NoStop}%
\bibitem [{\citenamefont {{Torres}}(2002)}]{51}%
  \BibitemOpen
  \bibfield  {author} {\bibinfo {author} {\bibfnamefont {D.~F.}\ \bibnamefont
  {{Torres}}},\ }\href {\doibase 10.1016/S0550-3213(02)00038-X} {\bibfield
  {journal} {\bibinfo  {journal} {Nuclear Physics B}\ }\textbf {\bibinfo
  {volume} {626}},\ \bibinfo {pages} {377} (\bibinfo {year} {2002})},\ \Eprint
  {http://arxiv.org/abs/hep-ph/0201154} {arXiv:hep-ph/0201154 [hep-ph]}
  \BibitemShut {NoStop}%
\bibitem [{\citenamefont {{Babichev}}\ \emph {et~al.}(2005)\citenamefont
  {{Babichev}}, \citenamefont {{Dokuchaev}},\ and\ \citenamefont
  {{Eroshenko}}}]{52}%
  \BibitemOpen
  \bibfield  {author} {\bibinfo {author} {\bibfnamefont {E.~O.}\ \bibnamefont
  {{Babichev}}}, \bibinfo {author} {\bibfnamefont {V.~I.}\ \bibnamefont
  {{Dokuchaev}}}, \ and\ \bibinfo {author} {\bibfnamefont {Y.~N.}\ \bibnamefont
  {{Eroshenko}}},\ }\href {\doibase 10.1134/1.1901765} {\bibfield  {journal}
  {\bibinfo  {journal} {J. Exp. Theor. Phys.}\ }\textbf {\bibinfo {volume}
  {100}},\ \bibinfo {pages} {528} (\bibinfo {year} {2005})},\ \Eprint
  {http://arxiv.org/abs/astro-ph/0505618} {arXiv:astro-ph/0505618 [astro-ph]}
  \BibitemShut {NoStop}%
\bibitem [{\citenamefont {{Babichev}}\ \emph {et~al.}(2013)\citenamefont
  {{Babichev}}, \citenamefont {{Dokuchaev}},\ and\ \citenamefont
  {{Eroshenko}}}]{53}%
  \BibitemOpen
  \bibfield  {author} {\bibinfo {author} {\bibfnamefont {E.~O.}\ \bibnamefont
  {{Babichev}}}, \bibinfo {author} {\bibfnamefont {V.~I.}\ \bibnamefont
  {{Dokuchaev}}}, \ and\ \bibinfo {author} {\bibfnamefont {Y.~N.}\ \bibnamefont
  {{Eroshenko}}},\ }\href {\doibase 10.3367/UFNe.0183.201312a.1257} {\bibfield
  {journal} {\bibinfo  {journal} {Physics Uspekhi}\ }\textbf {\bibinfo {volume}
  {56}},\ \bibinfo {eid} {1155-1175} (\bibinfo {year} {2013})},\ \Eprint
  {http://arxiv.org/abs/1406.0841} {arXiv:1406.0841 [gr-qc]} \BibitemShut
  {NoStop}%
\bibitem [{\citenamefont {{Ficek}}(2015)}]{54}%
  \BibitemOpen
  \bibfield  {author} {\bibinfo {author} {\bibfnamefont {F.}~\bibnamefont
  {{Ficek}}},\ }\href {\doibase 10.1088/0264-9381/32/23/235008} {\bibfield
  {journal} {\bibinfo  {journal} {Classical and Quantum Gravity}\ }\textbf
  {\bibinfo {volume} {32}},\ \bibinfo {eid} {235008} (\bibinfo {year}
  {2015})}\BibitemShut {NoStop}%
\end{thebibliography}%

\end{document}